\documentclass[12pt]{article}
\pdfoutput=1

\usepackage{amsmath}
\usepackage{amsfonts}
\usepackage{amssymb}
\usepackage{float} 
\usepackage{hyperref}
\newlength{\abstractwidth}
\usepackage{epsf}
\usepackage{color}

\usepackage{graphicx}

\renewcommand{\thefootnote}{\fnsymbol{footnote}}
\renewcommand{\thanks}[1]{\footnote{#1}}
\newcommand{\starttext}{
\setcounter{footnote}{0}
\renewcommand{\thefootnote}{\arabic{footnote}}}

\newcommand{\bea}{\begin{eqnarray}}
\newcommand{\eea}{\end{eqnarray}}
\newcommand{\ee}{\end{equation}}
\newcommand{\be}{\begin{equation}}

\newcommand{\no}{\nonumber}

\def\cI{{\cal I}}

\def\cN{{\cal N}}
\def\cO{{\cal O}}

\def\bC{{\bf C}}

\def\Re{{\rm Re}}
\def\Im{{\rm Im}}

\def\half{ {1\over 2}}
\def\p{\partial}

\def\a{\alpha}
\def\b{\beta}

\def\tet{\vartheta}

\def\o{\omega}

\def\f{\varphi}
\def\G{\Gamma}

\def\no{\nonumber}
\def\sm{\smallskip}

\long\def\symbolfootnote[#1]#2{\begingroup%
\def\thefootnote{\fnsymbol{footnote}}\footnote[#1]{#2}\endgroup}

\begin{document}

\starttext
\setcounter{footnote}{0}

\

\bigskip

\begin{center}

{\bf \Large Open  Worldsheets for  Holographic  Interfaces \footnote{This work was
supported in part by NSF grant PHY-07-57702.}}

\vskip 1in

{\large  Marco Chiodaroli, Eric D'Hoker and  Michael Gutperle }

\vskip .2in

{\sl Department of Physics and Astronomy }\\
{\sl University of California, Los Angeles, CA 90095, USA}\\
{\tt \small mchiodar@ucla.edu; dhoker@physics.ucla.edu; gutperle@physics.ucla.edu}

\vskip .7in

\begin{abstract}

\vskip 0.1in

Type IIB supergravity admits Janus and multi-Janus solutions with eight unbroken supersymmetries 
that are locally asymptotic to $AdS_3 \times S^3 \times M_4$ (where $M_4$ is either $T^4$ 
or $K_3$). These solutions are dual to two or more CFTs defined on half-planes which share a common 
line interface. Their geometry consists of an $AdS_2 \times S^2\times M_4$ fibration over
a simply connected Riemann surface $\Sigma$ with boundary. \\

In the present  paper, we show that regular exact solutions exist also for surfaces $\Sigma$ 
which are  not simply connected. Specifically, we construct in detail solutions for which $\Sigma$ has the 
topology of an annulus. This construction is generalized
to produce solutions for any surface $\Sigma$ with the topology of an open 
string worldsheet with $g$ holes. 

\end{abstract}

\end{center}

\newpage

\baselineskip=16pt
\setcounter{equation}{0}
\setcounter{footnote}{0}

\tableofcontents

\newpage

\baselineskip 16pt 

\section{Introduction}
\setcounter{equation}{0}

One of the best studied  examples of the AdS/CFT correspondence  \cite{Maldacena:1997re,Gubser:1998bc,Witten:1998qj} (for reviews see e.g. \cite{Aharony:1999ti,D'Hoker:2002aw}) is the duality 
between $AdS_3\times S^3 \times M_4$ vacua of type IIB string theory (where $M_4$
is either $T^4$ or $K_3$) and certain  
two-dimensional superconformal field theories with 16 supersymmetries
\cite{Witten:1997yu,Seiberg:1999xz,Vafa:1995bm,Dijkgraaf:1998gf}.

\sm

Deformations of the 
supersymmetric vacua which preserve half of the supersymmetries are of particular interest.  On the CFT 
side, such deformations can be associated with local operators as well as 
certain extended objects, such as Wilson loops, interfaces and surface operators. 
Defects and interfaces that preserve half of the conformal supersymmetry represent an important example of such
 extended objects for two-dimensional CFTs, and will be the focus of the present paper.

\sm

There are several methods for obtaining  spacetimes  that are dual to 
superconformal interfaces or defects.  
First, we can consider a brane intersection where a lower dimensional brane realizes a CFT on its
worldvolume while a higher dimensional brane introduces a conformal boundary \cite{Karch:2000gx}. In particular, one can add 
a D3 brane to a D1/D5 brane configuration introducing
a $0+1$-dimensional intersection. 
Without the additional D3 brane, the D1/D5 branes  produce the $AdS_3\times S^3\times M_4$ vacuum in the near-horizon limit.  
It has been argued \cite{Karch:2000gx} that the D3 brane survives this limit and introduces a defect in the dual theory.

Second, probe branes which have an $AdS_2$ worldvolume inside the $AdS_3$ space provide 
a holographic realization of one-dimensional conformal interfaces and defects
 \cite{Karch:2000gx,Aharony:2003qf,Bachas:2002nz,Bachas:2008jd,Raeymaekers:2006np,Yamaguchi:2003ay,Raju:2007uj,Mandal:2007ug}. 
In general, $\kappa$-symmetry of the worldvolume theory 
may be used to count the number of supersymmetries preserved by the probe
and to fix the positions of the probe branes in $AdS_3\times S^3\times M_4$ \cite{Skenderis:2002vf}.

Third, in many cases the extended supersymmetry of the interface or defect may be 
exploited to obtain supergravity solutions in analytic form. It is in this spirit that various
 regular Janus solutions with 16 supersymmetries were derived in type IIB supergravity 
\cite{D'Hoker:2006uv,D'Hoker:2007xy,D'Hoker:2007xz,D'Hoker:2007fq} and M-theory 
\cite{D'Hoker:2008wc,D'Hoker:2008qm,D'Hoker:2009my,D'Hoker:2009gg} 
to obtain holographic duals of interfaces, defects,  Wilson loops and surface operators
for super Yang-Mills theories in 2+1 and 3+1 dimensions.\footnote{For 
closely related work, see also \cite{Clark:2005te,Yamaguchi:2006te,Gomis:2006sb,Lunin:2006xr,Lunin:2007ab}.}

  Conformal superalgebras \cite{VanProeyen:1986me} 
provide a framework in which 
the above three methods may be understood in a unified way \cite{D'Hoker:2008ix}.  
In this paper, we will focus on the third approach and  derive exact supergravity solutions dual to interfaces and defects in 
$1+1$-dimensional CFTs.

\sm
Exact half-BPS solutions  in type IIB 
supergravity that preserve eight of the 16 supersymmetries of the 
$AdS_3\times S^3 \times M_4$ vacuum and are locally asymptotic to  
the vacuum solution were constructed in a previous paper \cite{Chiodaroli:2009yw} 
(see also \cite{Kumar:2002wc,Kumar:2003xi,Kumar:2004me,Bak:2007jm} for earlier related work). 
 The Ansatz for these solutions preserves 
a $SO(2,1)\times SO(3)$ subgroup of the full $SO(2,2)\times SO(4)$ isometry group 
of the vacuum. This $SO(2,1)\times SO(3)$ symmetry correctly reflects the global 
bosonic symmetry for a one-dimensional conformal interface, and uniquely extends 
to the superalgebra $SU(1,1|2)$ which has eight supersymmetries.

\sm

The exact solutions of \cite{Chiodaroli:2009yw} display a rich and interesting moduli space, and have two or more  asymptotic 
$AdS_3\times S^3 \times M_4$ regions which may be identified,
in the dual CFT,  with two-dimensional half-spaces glued together 
at the one-dimensional interface. In the different asymptotic regions, the dilaton 
and axion fields approach different constant values, and  the D1, D5, NS5 
and F1 charges take on different values. 

\sm
The goal of the present paper is to address a number of important questions 
which were raised in  \cite{Chiodaroli:2009yw}, but were not 
answered there:

\begin{itemize}
\item  The solutions of \cite{Chiodaroli:2009yw} have non-zero D1, D5, NS5, 
and F1 charges, but all have vanishing D3-brane 
charge. It would be interesting to find solutions carrying non-trivial D3-brane charge.
\item The solutions of \cite{Chiodaroli:2009yw} have spacetimes given by an 
 $AdS_2 \times S^2 \times M_4$ fibration over a Riemann surface $\Sigma$
with boundary. Moreover, $\Sigma$ is simply-connected, and may be conformally 
mapped to the upper half-plane. 
A similar analysis was conducted for half-BPS multi-Janus solutions dual to  $2+1$-dimensional
 interfaces and defects \cite{D'Hoker:2007xy,D'Hoker:2007xz}. In the latter case, 
the space-time is given by the warped product $AdS_4 \times S^2 \times S^2 \times \Sigma$, 
and regularity appears to require that $\Sigma$ has only a single boundary component and is simply-connected
though no rigorous {\sl theorem} to that effect 
is yet available.
For the lower-dimensional case of a one-dimensional interface or defect considered here,
however, this question must be re-examined. We intend to generalize the solutions 
of  \cite{Chiodaroli:2009yw} to the case where $\Sigma$  has a more complicated 
topology, involving more than one boundary component. 
\item The half-BPS solutions of \cite{Chiodaroli:2009yw} are completely regular,
and have the same symmetries and charges  associated with probe branes 
in the $AdS_3\times S^3$ vacuum. They may be viewed as the fully back-reacted 
solutions where the probe branes have been replaced by geometry and flux.  
A final question is whether the localized probe branes may be recovered by 
taking a (possibly singular) limit of the fully back-reacted solution. This reverse
process was possible in the original multi-Janus solutions 
\cite{D'Hoker:2007xy,D'Hoker:2007xz}, and is expected to be available here as well.
 \end{itemize}
In the present paper, we  show that the answers to the first two questions are intimately related. We  find that regular type IIB supergravity solutions for which $\Sigma$
has multiple boundary components do exist, and we  
construct them explicitly. For $\Sigma$ with the topology of an annulus (i.e. two 
boundary components), the construction is carried out explicitly in terms of 
elliptic functions and their related  Jacobi theta functions. For $\Sigma$ with the 
topology of a sphere with $g +1$ holes, with $g \geq 2$,   
the construction will be given in terms of the higher genus prime forms 
and theta functions of the double cover of $\Sigma$.

\sm

The question as to whether regular solutions for which 
$\Sigma$ also has handles can be obtained  will not be addressed in this paper,
though many of the tools needed to examine this question will be developed here.

\sm

Finally,  the new half-BPS interface and defect solutions we obtain  
will be an excellent laboratory for considering the probe limit of regular back-reacted solutions.

\subsection{Organization}

The structure of this paper is as follows.  In section  \ref{sec2} we briefly review the 
local half-BPS interface solutions  as well as the conditions imposed by global regularity. 
For more details  and the full derivations we refer the reader to  \cite{Chiodaroli:2009yw}.\\
In section \ref{sec3}, we consider the case of a Riemann surface with two disconnected 
boundary components (i.e. the annulus). We construct the solutions using theta functions,  
and solve the constraints imposed by regularity. We show that the solutions carry 
non-zero D3-brane charge, and that this charge is associated with the non-contractible 
cycle of the annulus. \\
In section \ref{sec4}, we examine the degeneration of the annulus 
where one boundary shrinks to a point, and show that, in this limit, extra asymptotic regions
 with $AdS_2 \times S^2 \times S^1 \times R$ geometry appear.\\
In section \ref{sec5}, we generalize the annulus solution to the case of a Riemann surface with an arbitrary number 
of boundary components utilizing the doubling trick to construct the solutions in terms of holomorphic differentials and 
 prime forms on the double Riemann surface. \\  
In the final section,  we   discuss possible generalizations of the solutions found in this paper and list several open questions 
and directions for future research. \\

\newpage

\section{Local half-BPS interface solutions}
\label{sec2}
\setcounter{equation}{0}

In this section, we present a summary of the Ansatz and local half-BPS solutions obtained in 
\cite{Chiodaroli:2009yw} which will be used again here. We also review the regularity and 
boundary conditions needed to promote the local results to globally well-defined solutions. 

\subsection{Ansatz}

In the local half-BPS solutions, the spacetime is constructed as a fibration of 
the product $AdS_2\times S^2\times M_4$ (where $M_4$ is either $T^4$ or $K_3$) 
over a two-dimensional Riemann surface $\Sigma$ with boundary. This product space
is invariant under the global symmetry group $SO(1,2)\times SO(3)$, and the appropriate
Ansatz must reflect this symmetry. The metric is given by,
\bea
ds^{2} = f_{1}^{2 } ds^{2}_{AdS_{2}} + f^{2}_{2}ds^{2}_{S^{2}} 
+ f^{2}_{3}ds^{2}_{M_{4}}  + \rho^{2 }dz  d\bar z
\eea
Symmetry requires that all  reduced bosonic fields, such as $f_1, f_2, f_3$, and $\rho$, 
depend only on $\Sigma$.  It will be convenient to introduce an orthonormal frame
associated with this metric; its components satisfy, 
\bea 
\eta_{i_{1}i_{2}}\; e^{i_{1}} \otimes e^{i_{2}} =  f^2_1  ds^{2}_{AdS_{2}} &\quad& i_{1,2}=0,1  
\no\\
\delta_{j_{1}j_{2}}\; e^{j_{1}} \otimes  e^{j_{2}} = f^2_2 ds^{2}_{S^{2}} &\quad& j_{1,2}=2,3 
\no \\ 
 \delta_{k_{1}k_{2}}\; e^{k_{1}} \otimes  e^{k_{2}} = f^2_3 ds^{2}_{M_{4}} &\quad& 
 k_{1,2}=4,5,6,7  
 \no \\
 \delta_{ab}\; e^{a} \otimes e^{b} =  \rho^{2 }dz d\bar z &\quad& a,b=8,9 
 \eea 
The standard one-form $P$ and composite $U(1)$ connection $Q$ may be expressed in 
terms of the dilaton $\Phi$ and axion $\chi$ field, as follows,
\bea
\label{compform}
 P= - d \Phi + {i\over 2} e^{- 2 \Phi} d\chi \hskip 1in  
 Q= - {1\over 2} e^{- 2 \Phi} d\chi 
\eea
Finally, the $SO(1,2)\times SO(3)$ symmetry restricts the three-form $G$ 
and five-form $F_5$ to be given by,
\bea
G & = & g^{(1)}_{a}e^{a01}+ g^{(2)}_{a}e^{a23}
\no \\
F_{5} &= & h_{a}  e^{a0123}+ \tilde h_{a} e^{a4567}
\eea
Self-duality of $F_5$ imposes the condition $h_{a}=-\epsilon_{a}^{\; \;b} \tilde h_{b}$. 
In this Ansatz, only the volume form on $M_4$ is taken into account,  while
the other cohomology generators of $M_4$ are omitted, and the corresponding 
moduli of these spaces are turned off. This explains why the only dependence 
on $M_4$ in the Ansatz for $G$ and $F_5$  is through the volume form $e^{4567}$.

\subsection{Local Solutions}

In \cite{Chiodaroli:2009yw},  the BPS equations and Bianchi identities were reduced 
to a system of four differential equations which admits a local solution
in terms of two harmonic functions,\footnote{To improve notational clarity, 
we denote here by $K$ the function that was denoted by $\hat h$ in \cite{Chiodaroli:2009yw}.}, $H$ and $K$,  and two holomorphic functions, $A$ and $B$. All 
supergravity fields of the local solution can be expressed in terms of these functions. 
The dilaton and axion are given by,
\bea
e^{4\Phi} & = & {1\over 4K^2} \Big( (A + \bar A)K  - (B + \bar B)^2 \Big) 
		\Big( (A + \bar A)K - (B - \bar B)^2  \Big) \\
 \chi &=& {i\over 2K} \Big( (A - \bar A)K -B^2 +\bar B^2  \Big) \label{axionform}
\eea  
The metric factors take the following form,
\bea
f^2_1 &=&  {c e^{- 2 \Phi} \over 2 f_3^2} {|H| \over K}  \Big( (A + \bar A) K -  (B - \bar B)^2 \Big) \label{sol-f1} \\
f^2_2 &=&  {ce^{- 2 \Phi} \over 2 f_3^2} {|H| \over K } \Big(  (A+ \bar A) K -   (B  + \bar B)^2  \Big) \label{sol-f2} \\
f_3^4 &=& 4  { c^2 e^{2\Phi} K \over A + \bar A} \label{sol-f3}
\eea
The constant $c$ is related to the volume of $K_3$ and was set to 1 in \cite{Chiodaroli:2009yw}.
The metric on $\Sigma$ can then be written as 
\bea
\rho^4= e^{2\Phi}K {|\partial_w H|^4  \over H^2} { A + \bar A \over |B|^4 } \label{sol-rho}
\eea
The following combinations of three-form fluxes and metric factors can be expressed as total derivatives, 
\bea  
f_2^2 \rho e^{-\Phi} \Re(g^{(2)})_z  &=&  \partial_w b^{(2)},\nonumber \\
 f_2^2 \rho e^{\Phi} \Im(g^{(2)})_z + \chi f_2^2 \rho e^{-\Phi} \Re(g^{(2)})_z &=&  \partial_w c^{(2)} \label{potdef2}
 \eea
There are analogous expressions for $g^{(1)}_z$ which can be found in \cite{Chiodaroli:2009yw} and that will not be needed in this paper.
The potentials written in terms of our holomorphic and harmonic functions are
\bea 
 b^{(2)} = -i  {H (B - \bar B) \over (A + \bar A) K- (B - \bar B)^2 } + \tilde h_1, 
&&  \tilde h_1={1 \over 2 i} \int {\partial_w H \over B} + c.c. 
 \label{potharmonic2app}
\\
c^{(2)} = - {H (A \bar B +  \bar A B) \over (A + \bar A) K - (B - \bar B)^2 } + h_2, 
&&  
h_2={1 \over 2 } \int {A \over B}\partial_w H + c.c.  
\label{potharmonic4app} 
\eea
Similarly, a combination of $F_5$ and metric factors can be written as a total derivative,
\bea \label{ckdef}
f^4_3 \rho \tilde h_z = \partial_w C_K, 
\hskip 0.7in
 C_K = {c^2 \over 2 i} {B^2 - \bar B^2 \over A + \bar A} - {c^2 \over 2} \tilde {K}  \label{fourfcharge}
\eea
where $\tilde h_1, \tilde h_2, \tilde K $ denote the  harmonic functions conjugate  to  
$h_1, h_2, K$ respectively.\\
Finally, we note that it is possible to rescale our functions by a constant $a$,
\be    K \rightarrow a^2 K, \qquad B \rightarrow a B, \qquad H \rightarrow a H, \qquad c \rightarrow a^{-1} c  \label{rescalesym}\ee
leaving all the physical fields invariant. 

\subsection{Regularity conditions}

Any choice of holomorphic functions $A,B$ and harmonic functions $H, K$ will 
produce bosonic fields which solve the type IIB supergravity field equations  and preserve 
eight real supersymmetries. In general, however,  such solutions may either 
have singularities or be unphysical, e.g. when a real scalar field 
like the dilaton $\Phi$ becomes complex. In order to guarantee sensible regular 
solutions, several
additional conditions have to be imposed. One such constraint comes from the fact 
that the asymptotic regions of the spacetime correspond locally to 
$AdS_3\times S^3\times M_4$.  The complete list of conditions to be satisfied is as follows:

\begin{itemize}
\item 
The radius of the $AdS_2$ slice, given by the metric coefficient $f_1$, 
is non-zero and finite everywhere, except at isolated singular points on the boundary 
of $\Sigma$.  Each such singular point corresponds to an $AdS_3 \times S^3\times M_4$ asymptotic region.
\item 
The radius of the $S^2$ slice, given by $f_2$, is finite in the interior of $\Sigma$,
and zero on the boundary of $\Sigma$. The boundary may be defined 
as the curve on which $f_2$ vanishes.
\item 
The radius of the $M_4$ manifold, given by $f_3$, and the dilaton combination 
$e^{2\Phi}$ are finite and non-zero everywhere on $\Sigma$, including the boundary.   
\end{itemize}
Using the above requirements, it was shown in \cite{Chiodaroli:2009yw} that the 
harmonic functions $H$, $K$, $A+ \bar A$ and $B + \bar B$ must obey vanishing 
Dirichlet boundary conditions, while the harmonic functions $i(A-\bar A)$, $i(B-\bar B)$ and $ \tilde K$ must obey Neumann  boundary conditions.
Moreover, $A$, $B$ and the meromorphic part of $K$ can admit only simple poles, and the following 
regularity conditions need to be respected,  

\begin{description}
\item[\;\;\;{\bf R1:}] 
All singularities of the harmonic functions $A+ \bar A$, $B + \bar B$ and $K$ must be common, the residues of $A$, $B$ and 
the meromorphic part of $K$ are related by $r_A r_K = r^2_B$ ;
\item[\;\;\;{\bf R2:}] 
The functions $A,B,H,K$ must be regular in the interior of $\Sigma$; 
\item[\;\;\;{\bf R3:}] 
The functions $A + \bar A $, $ K$ and $H$ cannot vanish in the interior  of $\Sigma$;
\item[\;\;\;{\bf R4:}] 
All the zeros of $B$ and $\partial_w H$ must be common.
\end{description}
Finally, there is an extra condition coming from the requirement that the dilaton field must be real or, more practically, that $e^{4 \Phi}$ must be positive,
\begin{description}
 \item[\;\;\;{\bf R5:}]  
The following inequality 
\bea 
 (A+ \bar A) K - (B + \bar B)^2 > 0 
 \label{reg-ineq} 
 \eea
must be obeyed throughout $\Sigma$, including on the boundary.
\end{description}
For a Riemann surface with a single boundary component, this follows as soon as 
conditions {\bf R1} to  {\bf R4} are satisfied. Similarly, we shall show that this condition does not pose 
any further constraint on the solutions constructed in this paper. 

\bigskip

\bigskip

\section{Two boundary components: the annulus}
\label{sec3}
\setcounter{equation}{0}

In \cite{Chiodaroli:2009yw} the Riemann surface $\Sigma$ had a single boundary 
component. The next simplest choice is a Riemann surface with two boundary 
components and no handles. By uniformization, it is always possible to map 
$\Sigma$ to the annulus, defined as the domain 
\bea
\label{annua}
\Sigma \equiv \left \{ w \in \bC, ~0 \leq \Re(w) \leq 1 , ~0  \leq \Im (w) \leq {t\over 2}  \right \}
\eea
where points $w+1 $ and $w$ are identified, giving indeed the topology of an 
annulus.  The two boundaries  $\partial \Sigma_{1,2}$ of the annulus are 
located at $\Im(w)=0$ and $\Im(w)=t/2$.

\sm

The annulus can be constructed from a Riemann surface without boundary, 
the so called double $\bar \Sigma$, which is a rectangular torus with pure imaginary modular  parameter $\tau=it$.
\bea
\label{torusa}
\bar \Sigma \equiv \{ w \in \bC, 0 \leq \Re(w) \leq 1 , 0 \leq \Im (w) \leq  t  \}
\eea
where $w+1$ is identified with $w$ and $w+i t$ is identified with $w$. 
The annulus (\ref{annua}) is obtained by quotient of  the  torus (\ref{torusa}) 
by an anti-conformal involution 
\bea
\Sigma = {\bar \Sigma } / \cI, \hskip 1in  \cI (z)={\bar z}
\eea
The boundaries $\partial \Sigma_{1,2}$ of the annulus are the fixed point set 
of the involution $\cI$. 

\subsection{Harmonic and holomorphic functions on the annulus}

The construction of the solution for the annulus proceeds in three steps. 
First, we construct a basic harmonic function with prescribed singularities 
and boundary conditions. Second, we express $A$, $H$ and $K$ using linear 
combinations of the basic function. Third, we find the meromorphic function $B$ 
which satisfies conditions {\bf R1} and {\bf R4}.

\begin{figure}
\centering
\includegraphics[ scale=0.50]{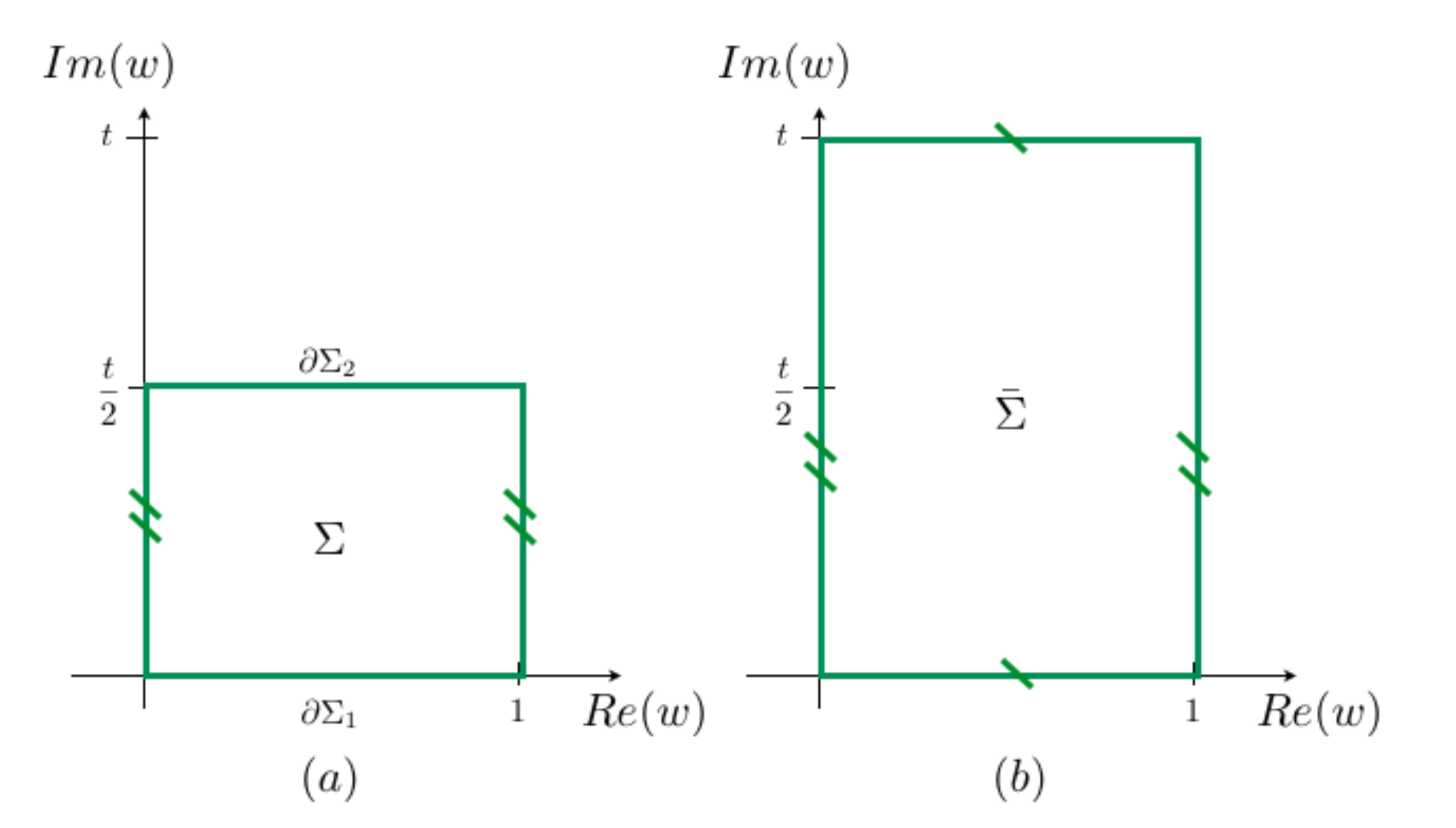}
\caption{(a) The annulus $\Sigma$ with boundaries at $\Im(w)=0, \Im(w)={t\over 2}$; 
(b) the doubled surface $\bar \Sigma $ is a torus with modulus $\tau=i t$. }
\label{annulus1}
\end{figure}

In the sequel,  we shall make use two Jacobi theta functions, which are defined 
as follows\footnote{We use conventions consistent with \cite{abramowitz}.}
\bea 
\vartheta_1 (u|\tau) &=& 2 \sum^{\infty}_{n=0} (-1)^{n} q^{(n+1/2)^2} \sin (2n+1)u 
\label{deftheta1} \\
\vartheta_4 (u|\tau) &=&1 + 2 \sum^{\infty}_{n=0} (-1)^{n} q^{n^2} \cos 2n u  
\label{deftheta4}
\eea
where $q=e^{ \pi i \tau }= e^{-\pi t}$. With these definitions, the quasi-periodicity conditions are,
\bea
 \vartheta_1 (u + \pi | \tau) & = & - \vartheta_1(u | \pi) 
 \no \\
  \vartheta_1(u+\pi \tau | \tau) & = & - e^{-2 i u} q^{-1} \vartheta_1(u |\tau) 
  \label{period-theta1} 
\eea
Furthermore under shifts by $\pi \tau/2$ we have,
\bea
\label{transfhalf}
\vartheta_1(u+ \pi \tau/2 |\tau) =   i   q^{-{1\over 4}} e^{- i u} \vartheta_4( u|\tau)
\eea
Note that $\vartheta_1$ is an odd function of $u$ and vanishes linearly as $u \to 0$,
while $\tet_4$ is an even function of $u$ which vanishes at $\pi \tau /2$.

\subsection{Construction of the basic harmonic function}

It turns out that the harmonic functions $A\pm \bar A,B \pm \bar B,H$, and $K$ 
may be simply constructed by linear superposition of a basic harmonic function, 
which has simple properties. The basic harmonic function $h_0$ on the  annulus 
$\Sigma$   satisfies  the following properties:\footnote{By abuse of notation, we 
shall refer interchangeably to the poles of the harmonic function $\mu(w)+\overline{\mu(w)}$
and the poles of its meromorphic part $\mu(w)$ only.}
\begin{enumerate}
\item 
$h_0$ has a single simple pole on  $ \p \Sigma$, so that the $(1,0)$-form 
$\p h_0$ has one double pole;
\item  
Away from the pole on $\p \Sigma$, $h_0$ satisfies Dirichlet 
conditions, i.e. $h_0=0$ on $\p \Sigma$.
\item 
$h_0 >0$ in the interior of $\Sigma$;
\end{enumerate}
The harmonic function $h_0$  can be expressed in terms of Jacobi theta 
functions,

\bea
\label{basicharma}
h_0(w, \bar w) =
i \left (  {\partial_w \vartheta_1 ({\pi w  }| \tau ) \over  \vartheta_1 
( {\pi w  }| {\tau} ) }+  {2\pi i w \over \tau} \right ) + c.c 
\eea
or as an infinite series of images,
\bea
\label{basicharma1}
h_0 (w, \bar w) = { \pi i \over \tau} \sin \left ( {\pi \over \tau}(w-\bar w) \right )
\sum _{m=-\infty} ^{+\infty} { 1 \over \left | \sin \left ( { \pi \over \tau} (w + m) \right ) \right |^2 }
\eea
The meromorphic part of (\ref{basicharma})  has a simple pole on $\partial \Sigma_1$ 
located at $w=0$ and hence condition~1 is satisfied. Since the meromorphic part in 
(\ref{basicharma})  is real when $\Im(w)=0$, the harmonic function $h_0$ satisfies  
vanishing Dirichlet boundary conditions at the first boundary $\partial \Sigma_1$. 
Using the transformation property (\ref{transfhalf}) one shows that the meromorphic 
part in (\ref{basicharma}) is real at $\Im(w)=t/2$ and hence $h_0$ satisfies 
vanishing Dirichlet boundary conditions at the second boundary $\partial\Sigma_2$. 
Hence condition 2 is satisfied. This property is actually manifest from the prefactor in 
(\ref{basicharma1}). The harmonic function (\ref{basicharma}) is single 
valued on the annulus,
\bea
h_0 (w+1,\bar w+1) = h_0(w,\bar w)
\eea
and vanishes on both boundary components.
By the maximum principle for harmonic functions it follows that $h_0>0$ in the interior of the annulus\footnote{The harmonic function could also be strictly negative, however explicit evaluation of the harmonic function  at special points in the interior shows that it takes positive values.}. Hence condition 3 is also satisfied. 

\sm

The basic harmonic function $h_0$ has a singularity at $w=0$ at the first boundary 
component $\partial \Sigma_1$. The location of the singularity can be shifted to any 
point 
on the boundary by a real translation so that $h_0(w-x,\bar w-x)$ has a singularity at 
$w=x$. 
To obtain harmonic functions which have  singularities  on the second boundary component 
$\partial \Sigma_2$, we define
\bea
w' \equiv  {\tau \over 2}- w
\eea
and  for a real $y$, the harmonic function $h_0(w' + y,\bar w'+y)$ has a pole on the 
second boundary at $w= \tau /2 + y$ and satisfies conditions 1-3.

\sm

The  harmonic function $\tilde h_0$ conjugate to the basic harmonic function $h_0$ 
is  given by
\bea
\tilde h_0(w, \bar w) 
= \left (  {\partial_w \vartheta_1 ({\pi w  }| \tau ) \over  \vartheta_1 ({\pi w  }| {\tau}) }
+  {2\pi i w\over \tau} \right ) + c.c 
\eea
Note that, unlike $h_0$, the conjugate harmonic function $\tilde h_0$ is not 
single-valued on the annulus. Its monodromy is given as follows,
\bea
\label{mono1}
\tilde h_0(w+1, \bar w+1) =\tilde h_0(w, \bar w)+ {4\pi i \over \tau} 
\eea

\subsection{Construction of the supergravity solution on the annulus}

The harmonic functions $A\pm \bar A, B \pm \bar B, H$ and $K$ may now be expressed 
as linear combinations of basic harmonic functions with a pole at various points on both boundaries.  Note that for some of the harmonic functions, both the harmonic function and  
its harmonic conjugate must be single-valued on the torus. This condition imposes  
constraints on the coefficients of the linear combination. In this section we will construct 
the four harmonic functions which determine our solution.

\medskip

$\bullet$ The function $H$ must be single-valued and positive on $\Sigma$, and obey 
vanishing Dirichlet boundary conditions.  We take $H$ to have $N$ poles $x_{H,\a}$ with
$\a = 1, \cdots, N$ on $\p \Sigma _1$, and $N'$ poles $y_{H,\b}$ with 
$\b = 1,\cdots, N'$ on $\p \Sigma _2$.  The corresponding residues
$r_{H,\a}$ and $r'_{H,\beta}$ must be positive, but are otherwise left undetermined, 
\bea 
\label{Hform}
H = \sum^{N}_{\alpha=1} r_{H,\alpha} h_{0}(w-x_{H,\alpha}, \bar w - x_{H,\alpha})
+ \sum^{N'}_{\beta=1} r'_{H,\beta} h_{0}(w'+y_{H,\beta}, \bar w' + y_{H,\beta}) 
\eea
It is possible to use the translation symmetry of the annulus to fix the position of the first singularity at $x_{H,1} = 0$. Since the conjugate harmonic function $\tilde H$  does 
not appear in any of the expressions for the supergravity fields of the solution, $\tilde H$ 
does not need to be single-valued on the annulus. Consequently, there are  no additional conditions involving the residues and the number of parameters is equal to $2N + 2N'-1$. 
Note that each pole $x_{H,\alpha}$ and $y_{H,\beta}$ of $H$ corresponds to an 
asymptotic $AdS_3 \times S^3$ region. Hence, the full supergravity solution will have a total of $N+N'$ 
asymptotic  $AdS_3 \times S^3$ regions.

\sm

$\bullet$ The harmonic function $A + \bar A$ must be single-valued and positive in $\Sigma$,
and obey vanishing Dirichlet boundary conditions. The conjugate function
$-i(A - \bar A)$ must also be single-valued on $\Sigma$.  
As a result, these functions take on the following form,
\bea 
\label{Aform}
A + \bar A &=& \sum^{M}_{\alpha=1} r_{A,\alpha} h_0(w-x_{A,\alpha}, \bar w - x_{A,\alpha})+\sum^{M'}_{\beta=1} r'_{A,\beta} h_0(w' + y_{A,\beta}, \bar w' + y_{A,\beta})
\qquad \\
-i(A - \bar A) &=& \sum^{M}_{\alpha=1} r_{A,\alpha} \tilde h_0(w-x_{A,\alpha}, \bar w - x_{A,\alpha})+\sum^{M'}_{\beta=1} r'_{A,\beta} \tilde h_0(w'+y_{A,\beta}, \bar w' + y_{A,\beta})  
\eea 
Single-valuedness of $-i(A-\bar A)$ imposes a relation between the residues.
Using (\ref{mono1}), one obtains the monodromy of this function,
\bea 
-i(A - \bar A) (w+1,\bar w + 1) = -i(A - \bar A) (w,\bar w) 
+ {4 \pi i  \over \tau} \left (\sum^M_{\alpha=1} r_{A,\alpha}
- \sum_{\beta=1}^{M'} r'_{A,\beta} \right ) 
\eea
Hence, $-i(A-\bar A)$ will be single-valued provided the following relation is obeyed,
\bea  \label{asingleval}
\sum^M_{\alpha=1} r_{A,\alpha} = \sum_{\beta=1}^{M'} r'_{A,\beta} 
\eea
This condition brings the number of parameters in the definition of $A$ down to $2M+2M'$. Note that one of the parameters 
is given by the constant in the definition of the dual harmonic function $-i(A-\bar A)$.  

\sm

$\bullet$ The meromorphic function $B$ must have the same poles as $A$
(according to the regularity condition {\bf R1}), and the same zeros as $\partial_w H$ 
(according to {\bf R4}). Moreover, both $B+\bar B$ and $-i(B- \bar B)$ must be single-valued. 
The functions $B+\bar B$ and $-i(B-\bar B)$ are not subject, however, to any positivity 
condition, and are allowed to change sign inside $\Sigma$. We get the following 
expression for $B$:
\bea 
B(w) =  { \partial_w H (w) \over \mathcal{N}(w)} 
\eea
Here, $\cN(w) dw$ is a single-valued meromorphic form of weight $(1,0)$ whose zeros and poles
are determined by the requirement that $B$ and $\p_wH$ have common zeros
and $B$ and $A$ have common poles. Thus, $\cN$ must cancel the (double) poles
of $\p_wH$, and reproduce the poles of $A$. The resulting form may be expressed
in terms of the  Jacobi theta function $\tet_1$, 
\bea\label{Nform}
\mathcal{N}(w) 
= e^{  i \pi  \phi  w} ~  
{ \prod^M_{\alpha=1} \vartheta_1 \Big( \pi (w-x_{A,\alpha}) \Big. \Big| \tau \Big) 
\prod^{M'}_{\beta=1} \vartheta_1 \Big(\pi ( w'+y_{A,\beta}) \Big. \Big| \tau \Big) \over 
\prod^N_{\alpha=1} \vartheta_1 \Big( \pi (w-x_{H,\alpha}) \Big. \Big| \tau \Big)^2 
\prod^{N'}_{\beta=1} \vartheta_1 \Big(\pi ( w'+y_{H,\beta}) \Big. \Big| \tau \Big)^2 }  
\eea
Here, the exponential prefactor has been included in order to make $\cN$ 
properly single-valued, and $\phi$ is a real constant phase. Note that we can fix an overall real constant 
in the definition of $B$ using the symmetry (\ref{rescalesym}).

\sm

It remains to work out the precise conditions required to render $\cN$ single-valued.
Using the transformation property (\ref{period-theta1}) we see that single-valuedness of
$B$, namely $B(w+1)=B(w)$ requires $M+M'+\phi$ to be an even integer. 
Moreover, in order to have $B + \bar B$ obey vanishing Dirichlet boundary conditions,
we need $\mathcal{N}(w)$ to be real for $w=x$ and $w= \tau/2 + x$ with $x$ real. 
Using (\ref{transfhalf}) $\vartheta_1$  can be re-expressed in terms of $\vartheta_4$ when 
evaluated at $\partial \Sigma_2$,
\bea 
\vartheta_1 \Big( \pi { \tau \over 2} -\pi (x-y_{H,\beta} ) \Big. \Big| \tau \Big) &=& i q^{-1/4} e^{i \pi ( x-y_{H,\beta}) } \;  \vartheta_4 \Big( \pi ( y_{H,\beta}-x )\Big. \Big| \tau \Big) \label{halfperiod1} \\
 \vartheta_1 \Big(  \pi{\tau \over 2}  +\pi (x-x_{H,\alpha} )\Big. \Big| \tau \Big) &=& i q^{-1/4} e^{-i \pi (x-x_{H,\alpha} ) } \; \vartheta_4 \Big( \pi (x-x_{H,\alpha} )  \Big. \Big| \tau \Big) \label{halfperiod2}
 \eea
Both $\vartheta_1$ and $\vartheta_4$ are real when their argument is real and 
$\tau$ is purely imaginary, as we can see from equations (\ref{deftheta1}) 
and (\ref{deftheta4}). Using the above expressions, we see that the phases $\varphi$  of 
$\mathcal{N}(x)$ and $\mathcal{N}(\tau/2 + x)$ are respectively given by:
\bea \label{singlevalphi}
\varphi \Big (\mathcal{N}(x) \Big) &=& 
\pi  \Big( x (\phi - 2 N' + M') - \sum^{M'}_{\beta=1} y_{A,\beta} 
+ 2 \sum^{N'}_{\beta=1} y_{H,\beta} \Big) - M' {\pi \over 2 } 
\\
\varphi \Big (\mathcal{N}(x+ \tau/2) \Big) &=& 
\pi \Big( x (\phi + 2N - M ) - 2\sum^N_{\alpha=1} x_{H, \alpha}
 + \sum^M_{\alpha=1} x_{A,\alpha} \Big) - M {\pi \over 2}  
\eea
To cancel the $x$-dependence of these phases we need $\phi = M- 2N= 2N'-M'$.
Upon eliminating $\phi$, we have, 
\bea 
M+M' = 2(N+N') 
\label{cond-annulus1}
\eea
The earlier requirement that $\phi+M+M'$ is an even integer then implies that 
$2N+M'$ must be an even integer, so that both $M$ and $M'$ must be even integers.
Finally, in order to have real $\mathcal{N}(w)$ on the boundaries we need the 
$x$-independent parts of the phases to vanish as well, and we have, 
\bea
4 \sum^{N'}_{\beta=1} y_{H,\beta} 
- 2 \sum^{M'}_{\beta=1} y_{A,\beta}   & \equiv & 0 ~
 (\text{mod}\; 2) 
 \label{cond-annulus2} \\
4 \sum^N_{\alpha=1}  x_{H, \alpha}  
- 2 \sum^M_{\alpha=1} x_{A,\alpha}  & \equiv & 0 ~
 ( \text{mod} \; 2) 
 \label{cond-annulus3} \eea
If the above conditions are satisfied, then $B+\bar B$ obeys Dirichlet boundary conditions. 
Since $B+ \bar B$ is harmonic, single-valued and with the same poles of $A+\bar A$, 
it must be possible to express it directly in terms of the basic harmonic function $h_0$ as well,
\bea 
\label{Bform}
B + \bar B = \sum^{M}_{\alpha=1} r_{B,\alpha} h_0(w-x_{A,\alpha}, \bar w- x_{A,\alpha})+\sum^{M'}_{\beta=1} r'_{B,\beta} h_0(w' + y_{A,\beta}, \bar w' + y_{A,\beta}) 
\eea
for some residues $r_{B,\alpha}$ and $r'_{B, \beta}$, $\alpha= 1 \dots M$, $\beta= 1 \dots M'$. Note that these residues do not have necessarily 
the same sign.

\medskip

$ \bullet$ The harmonic function $K$ has the same poles as $A+ \bar A$ and the 
residues are given by the condition {\bf R1}.
The expression is fixed to be:
\bea
\label{Kform}
K = \sum^{M}_{\alpha=1} {r^2_{B,\alpha} \over r_{A, \alpha}} 
h_0(w-x_{A,\alpha}, \bar w - x_{A,\alpha})
+\sum^{M'}_{\beta=1} {r^{'2}_{B,\beta} \over r_{A, \beta}} 
h_0(w' + y_{A,\beta}, \bar w' + y_{A,\beta}) 
\eea
In this expression, $K$ is manifestly positive everywhere on $\Sigma$.
Note that the dual harmonic function $\tilde K$ does not need to be single-valued. So far, our solution is dependent on a total number of parameters
\be\label{noofpara}
 {\rm no.\; of\; parameters: \;\;\;} 2(N+N')+2(M+M') = 6(N+N')
\ee
In this counting  the modular parameter of the annulus  $\tau= i t$, the constant $c$ and a constant in the definition of $\tilde K$ are included.

\subsection{Regularity conditions \label{secR5}}

At this stage, the harmonic functions satisfy the  regularity conditions {\bf R1}, {\bf R2}, 
{\bf R3} and {\bf R4}. The last remaining condition is {\bf R5}, namely that 
\bea 
K  (A + \bar A) - (B + \bar B)^2 >0 
\eea
everywhere in $\Sigma$. We shall now show that the inequality follows 
from the form of the solution.

\sm

We begin by proving the $\geq$ inequality. To do so, we introduce the following abbreviations,
\bea
h_\a & \equiv & h_0(w-x_{A,\alpha}, \bar w - x_{A,\alpha}) 
\hskip 1in \a = 1, \ldots , M
\no \\
h_\b' & \equiv & h_0(w' + y_{A,\beta}, \bar w' + y_{A,\beta}) 
\hskip 1in \b = 1, \ldots , M'
\eea
For $w,w'$ in the interior of $\Sigma$, both quantities are strictly 
positive. The precise values taken by these functions will be immaterial
in the proof below. To evaluate $K  (A + \bar A) - (B + \bar B)^2$, we use
the expressions for $(A+\bar A)$, $(B+\bar B)$ and $K$ from (\ref{Aform}),
(\ref{Bform}) and (\ref{Kform}) respectively. 
Since the residues $r_{A,\a}$ and $r_{A,\b}' $ are strictly positive, 
we can define the following combinations,
\bea
v_\a = \left ( r_{A,\a} h_\a \right )^\half 
& \hskip 1in & 
u_\a = {r _{B,\a} h_\a \over v_\a} 
\no \\ 
v'_\b = \left ( r_{A,\b}' h_\b' \right )^\half 
& \hskip 1in &
u'_\b = {r _{B,\b}' h_\b' \over v_\b'} \label{def-uv}
\eea
In terms of these variables we have 
\be
K (A + \bar A) - (B + \bar B)^2
 = 
\left ( \sum _\a u_\a^2  + \sum _\b u_\b '^2 \right )
\left ( \sum _\a  v_\a^2   + \sum _\b  v_\b '^2 \right )
-\left ( \sum _\a  u_\a v_a   + \sum _\b  u_\b' v'_\b \right )^2 \no
\ee
The right hand side is automatically positive or zero in view of Schwartz's inequality, 
so that we have,
\bea
K (A + \bar A) - (B + \bar B)^2 \geq 0
\eea
Equality is obtained if and only if the $M+M'$ dimensional vectors $(u_\a, u_\b')$
and $(v_\a, v_\b')$ are proportional to one another, so that $u_\a = \lambda v_\a$
and $u_\b ' = \lambda v_\b '$ for some real number $\lambda$.
This proportionality relation implies that $r_{B\a} h_\a = \lambda r_{A\a} h_\a$ and $r_{B,\b}' h_\b ' = \lambda r_{A,\b}' h'_\b$,
as we can see from (\ref{def-uv}). We will rule out this possibility by showing that at least one of the residues $r_{B,\alpha}$, $r'_{B,\beta}$ has negative sign.\\
First, we note that $\partial_w H$ cannot vanish on the boundary of $\Sigma$. This follows from the fact that $H$ can be expanded 
close to any non-singular point $x$ on $\partial \Sigma_{1,2}$ as,
\be H = i \partial_w H (x) (w-x) + {i \over 2} \partial^2_w H(x) (w-x)^2 + \ldots + c.c. \ee
If $\partial_w H$ has a zero of order p for $w=x$, the above expansion can be rewritten as follows,
\be H = \text{const} \; \Im (w-x)^{p+1} + \ldots = \text{const} \; r^{p+1} \sin (p+1) \phi + \ldots \ee 
where we have introduced polar coordinates close to $w=x$. Because of the sine function, if $p>0$ then $H$ has some zeros in the bulk 
of $\Sigma$. However, this is not possible because $H$ is strictly positive in the interior of $\Sigma$ by construction.
Hence, the zeros of   $\partial_w H$ cannot be on $\partial \Sigma_{1,2}$ and must be located in the interior of $\Sigma$.\\
The condition {\bf R4} forces $B$ to have the same zeros as $\partial_w H$. 
Therefore, $B$ must vanish somewhere in the interior of $\Sigma$ 
and some of its residues must be negative.  This allows us to exclude that $r_{B\a} h_\a = \lambda r_{A\a} h_\a$ and $r_{B,\b}' h_\b ' = \lambda r_{A,\b}' h'_\b$
for all $\a$ and $\b$. Hence, our solutions satisfy {\bf R5} without any additional condition. \\ 

\subsection{Charges of the solutions}

Our solutions display $N+N'$ non-trivial three-spheres. In general, a three-sphere will correspond to a curve on $\Sigma$   
starting and ending on the boundary. 
We can construct a basis of three-spheres by choosing $N+N'-1$ curves having support in the neighborhood of 
singularities of $H$, so that each curve starts on the boundary on one side of the pole and ends  on the
opposite side of the same boundary. \\ 

\begin{figure}
\centering
\includegraphics[ scale=0.18, angle=270]{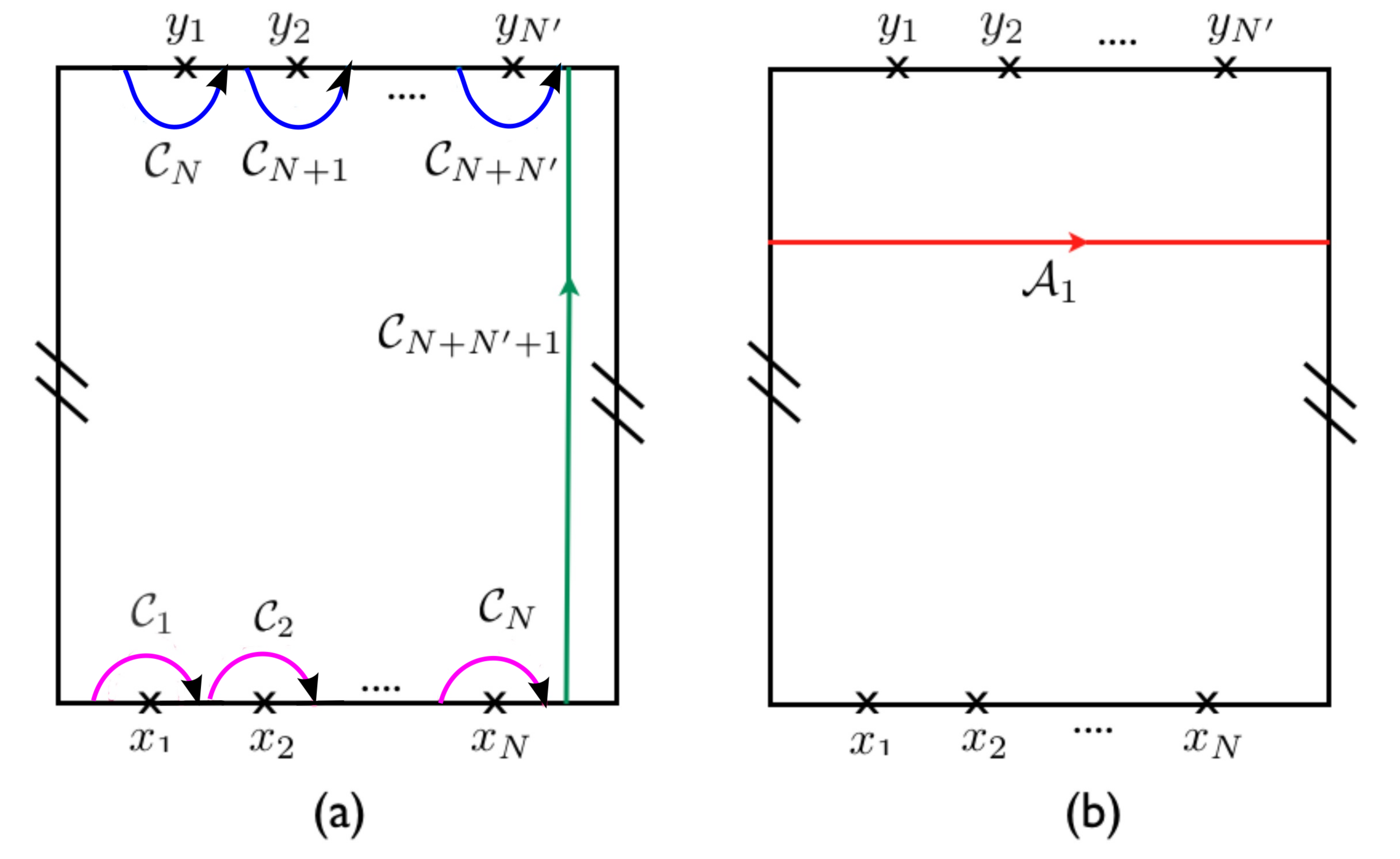}
\caption{(a) Choice of curves ${\cal C}_i$ which are associated with $N+N'$ linearly independent five-brane charges. (b) Non-contractible cycle ${\cal A}_1$ associated  with three-brane charge}
\label{cycle1}
\end{figure}

We choose a set of curves  ${\cal C}_i,\; i=\{ 1,2,\ldots N+N'+1\}$ as follows (see figure \ref{cycle1} (a)). 
The curves ${\cal C}_{\a},\; \{ \a=1,2,\ldots,N\}$ are surrounding the pole $x_{H,\alpha}$ on $\partial \Sigma_1$, the curves ${\cal C}_{N+\beta}, \;\{ \beta=1,\ldots,N'\} $
are surrounding the  poles $y_{H,\beta}$ on $\partial \Sigma_2$. The last curve $C_{N+N'+1}$ is chosen to  start on the lower boundary and end on the upper boundary.\\

\noindent The flux of the three-form field on each of the spheres gives the total enclosed NS5 charge,
\bea q_{NS5}({\cal C}) &=& {1 \over {\rm Vol}(S^3)} \int_{{\cal C} \times S^2} e^{- \Phi} Re(G) \nonumber \\
&=& {1 \over \pi} \int_{\mathcal C} f^2_2 e^{-\Phi} \rho  Re(g^{(2)})_z dw + {1 \over \pi} \int_{\mathcal C} f^2_2 e^{-\Phi} \rho Re(g^{(2)})_{\bar z} d\bar w
\no \\ &=& {1 \over \pi}  \int_{\mathcal C}  \big( \partial_w b^{(2)} dw+ \partial_{\bar w} b^{(2)} d\bar w\big)  \label{fiebrchg}  
\eea
Similarly, the  D5 Maxwell charge is given by \footnote{See \cite{Marolf:2000cb} for a discussion 
on the different notions of charges in presence of Chern-Simons terms.}
\bea
 q^M_{D5} ({\cal C})&=& {1 \over {\rm Vol}(S^3)} \int_{{\cal C} \times S^2} e^{\Phi} Im(G) \no \\
&=& {1 \over \pi}  \int_{\mathcal C}  \Big( \big( \partial_w c^{(2)} - \chi \partial_w b^{(2)} \big) dw + \big( \partial_{\bar w} c^{(2)} - \chi 
\partial_{\bar w} b^{(2)} \big) d\bar w \Big) \label{fivebranechMax}
\eea
The D5 Page charge has a similar expression,
\bea
 q^{P}_{D5} ({\cal C})&=& {1 \over {\rm Vol}(S^3)} \int_{{\cal C} \times S^2} \Big( e^{\Phi} Im(G)-\chi e^{- \Phi} Re(G) \Big) 
={1 \over \pi}  \int_{\mathcal C}  \big( \partial_w c^{(2)} dw+ \partial_{\bar w} c^{(2)} d\bar w\big) \qquad \label{fivebranechBS}
\eea
The expressions for $b^{(2)}$ and $c^{(2)}$ were defined in (\ref{potharmonic2app},\ref{potharmonic4app}), while $V_3$ is the volume of the unit three-sphere.
The value of the five-brane charges (\ref{fiebrchg}) and (\ref{fivebranechBS}) does not change upon deformation of the contours $C_i$ as long as one does 
not cross any of the singularities $x_{H,\alpha}$ and $y_{H,\beta}$.  Deforming the contours for the charges associated with the singularities one can show the following relations for the conservation of charges,
\bea
\sum_{i=1}^{N+N'} q^{P}_{D5}({\cal C}_i) =0, \quad  \sum_{i=1}^{N+N'} q_{NS5}({\cal C}_i) =0
\eea
Which shows that one of the charges associated with the singularities of $H$ is linearly dependent.

\medskip

The explicit expression for the charges associated with the asymptotic regions can be obtained by evaluating the integrals (\ref{fiebrchg})  for contours
which are contracted towards the singularity. For concreteness we consider the singularities on the lower boundary. 
The relevant function  can be expanded in as $w \to x_{H, \alpha}$,
\bea H &\sim& {2 y r_{H,\alpha} \over |w- x_{H,\alpha}|} \\
A &\sim& i a^{(0)}_\alpha+ i a^{(1)}_\alpha  (w- x_{H,\alpha}) \\
B &\sim& i b^{(0)}_\alpha+ i b^{(1)}_\alpha  (w- x_{H,\alpha}) \\
K &\sim& i k^{(0)}_\alpha+ i k^{(1)}_\alpha  (w- x_{H,\alpha}) + c.c. \eea   
The contribution to the integrals come from the functions $\tilde h_1$ and $h_2$ in (\ref{potharmonic2app},\ref{potharmonic4app}). We get the following expressions,
\bea 
\tilde h_1 & \sim & - { y\; r_{H,\alpha}   \over   b^{(0)}_\alpha|w-x_{H,\alpha}|^2  } +{1 \over 2} {r_{H,\alpha} b^{(1)}_\alpha\over (b^{(0)}_\alpha)^2} 
\log {w-x_{H,\alpha} \over \bar w - x_{H,\alpha}} \\
h_2 & \sim & - { y\; r_{H,\alpha}  a^{(0)}_\alpha  \over  b^{(0)}_\alpha |w-x_{H,\alpha}|^2  } +{r_{H,\alpha} \over 2}  
\Big( { a^{(0)}_\alpha b^{(1)}_\alpha\over (b^{(0)}_\alpha)^2  } - { a^{(1)}_\alpha  \over  b^{(0)}_\alpha } \Big) \log {w-x_{H,\alpha} \over \bar w - x_{H,\alpha}} \quad
\eea

Since the endpoints of the curves are taken on the boundaries, the terms proportional to $y$ give a zero contribution and we are left with the following expressions,
\be q_{NS5}({\cal C}_\alpha) = {r_{H,\alpha} b^{(1)}_\alpha\over (b^{(0)}_\alpha)^2 }, \quad 
q^{M}_{D5}({\cal C}_\alpha) = {r_{H,\alpha} \over b^{(0)}_\alpha}  \Big( { ( b^{(1)} _\alpha)^2 \over k^{(0)}_\alpha } -  a^{(1)}_\alpha \Big), \quad
q^{P}_{D5}({\cal C}_\alpha) = r_{H,\alpha}  \Big( { a^{(0)}_\alpha b^{(1)} _\alpha \over (b^{(0)}_\alpha)^2 } -{ a^{(1)}_\alpha \over b^{(0)}_\alpha }\Big) \qquad
\ee  
with analogous equations for the poles on the upper boundary. The expressions for the charges associated with $C_{N+N'+1}$  are given by the integrals
\bea
 q_{NS5}({\cal C}_{N+N'+1}) &=& {1\over 2i} \int_{C_{N+N'+1}} \Big({1\over B} \partial_w{H}  dw- {1\over \bar B} \partial_{\bar w}{H}  d\bar w \Big) 
= {1\over 2i} \int_{C_{N+N'+1}} \Big({\cal N}   dw- \bar{\cal N}    d\bar w\Big) \qquad \no \\
 q^P_{D5}({\cal C}_{N+N'+1})&=&{1\over 2} \int_{C_{N+N'+1}} \Big({A\over B} \partial_w{H}  dw+{\bar A\over \bar B} \partial_{\bar w}{H}  d\bar w\Big) 
= {1\over 2} \int_{C_{N+N'+1}}  \Big({A \cal N}  dw+\bar  A \bar{\cal N}    d\bar w\Big) \qquad \no \eea
These expressions provide a physical interpretation for the $(1,0)$-form ${\cal N}$ introduced in (\ref{Nform}). 
Note that these charges are not associated with any asymptotic region.\\
\medskip

The three-brane charge of the solution is given by the integral of the self-dual five form over $K_3\times {\cal C}$ where ${\cal C}$ is a curve in $\Sigma$.
\be
q_{D3}({\cal C})= {1\over {\rm Vol} (S^5)}\int_{K_3\times {\cal C}} F_5 = {3\over 4\pi^3} \int_{ {\cal C}} f^4_3 \rho \big( \tilde h_z  dw+  \tilde h_{\bar z} d\bar w\big)
= {3\over 4\pi^3}  \int_{ {\cal C}} \big( \partial_w C_K dw+  \partial_{\bar w} C_K d\bar w\big) \qquad
\ee
$C_K$ is defined in (\ref{ckdef}). 
Since the annulus has a non-contractible cycle ${\cal A}_1$ (see figure \ref{cycle1} (b)), the three-brane charge integrated over $K_3\times {\cal A}_1$ can be nonzero
 due to the fact that $\tilde K$ has nontrivial monodromy around ${\cal A}_1$.
\bea\label{dthreech}
q_{D3}({\cal A}_1)&=&  {3 c^2 \over  \pi^2 t }  \Big(
 \sum^{M}_{\alpha=1} {r^2_{B,\alpha} \over r_{A, \alpha}} 
-\sum^{M'}_{\beta=1} {r^{'2}_{B,\beta} \over r_{A, \beta}} \Big)  
\eea
(\ref{mono1})  and (\ref{Kform}) were used to derive the above equation. 

\subsection{Examples \label{example}}
In this section we present the solutions with the smallest number of parameters. The conditions which constrain the minimal solutions are the following.\\
First, equation (\ref{asingleval}), together with the positivity of the residues $ r_{A,\alpha}$ and $r'_{B,\beta}$, 
implies that $M$ and $M'$ both have to be nonzero. Second,  the conditions (\ref{singlevalphi}) imposed by the fact that $B$ has
to be single-valued, imply that both  $M$ and $M'$ must be even integers.\\
\begin{figure}[t]
\begin{center}
\includegraphics[ scale=0.3]{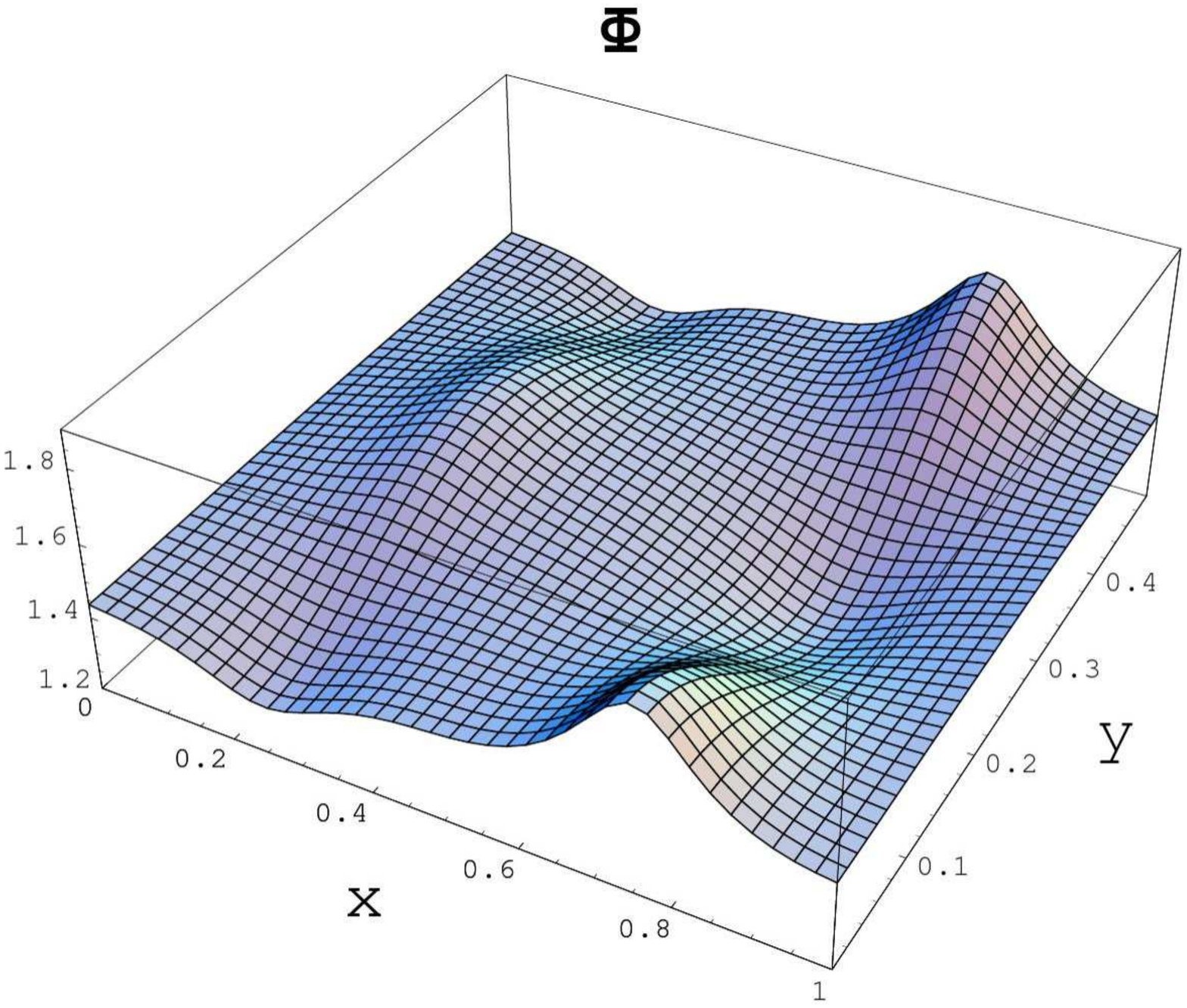}\includegraphics[ scale=0.3]{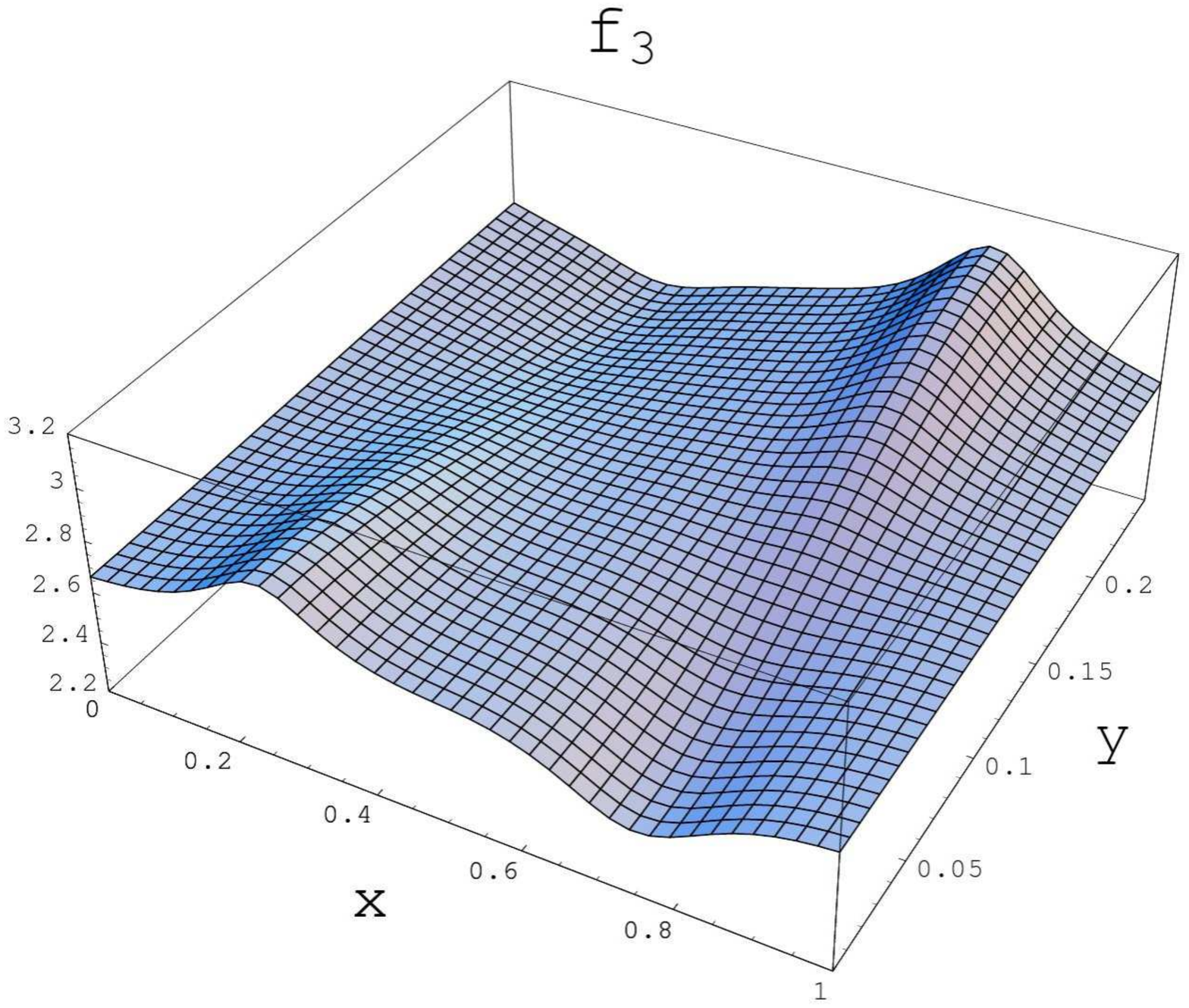}\\
\caption{Dilaton and metric factor $f_3$ for the example of section \ref{example}.}
\label{example1b}
\end{center}
\end{figure}
\begin{figure}[t]
\begin{center}
\includegraphics[ scale=0.3]{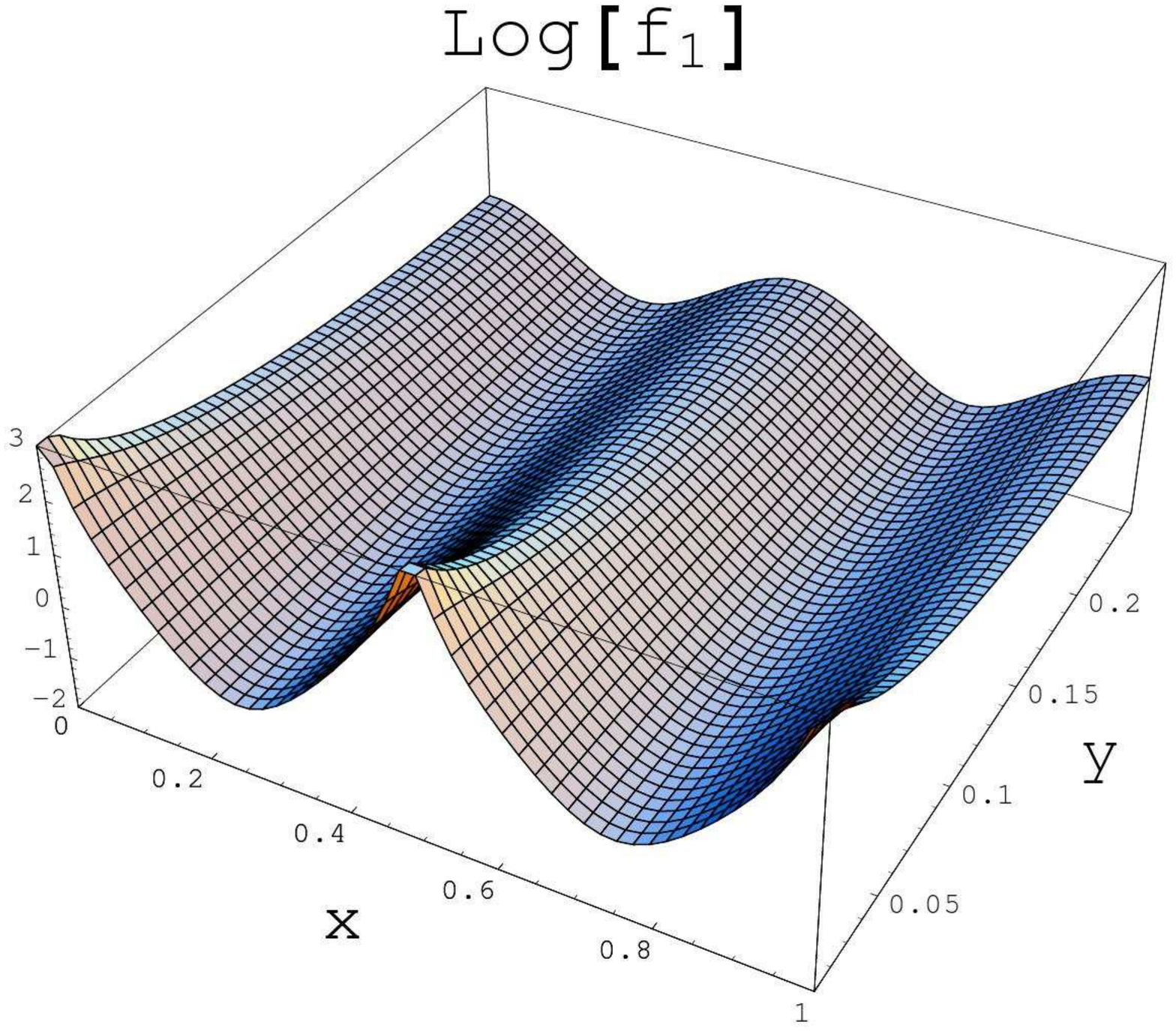}\includegraphics[ scale=0.3]{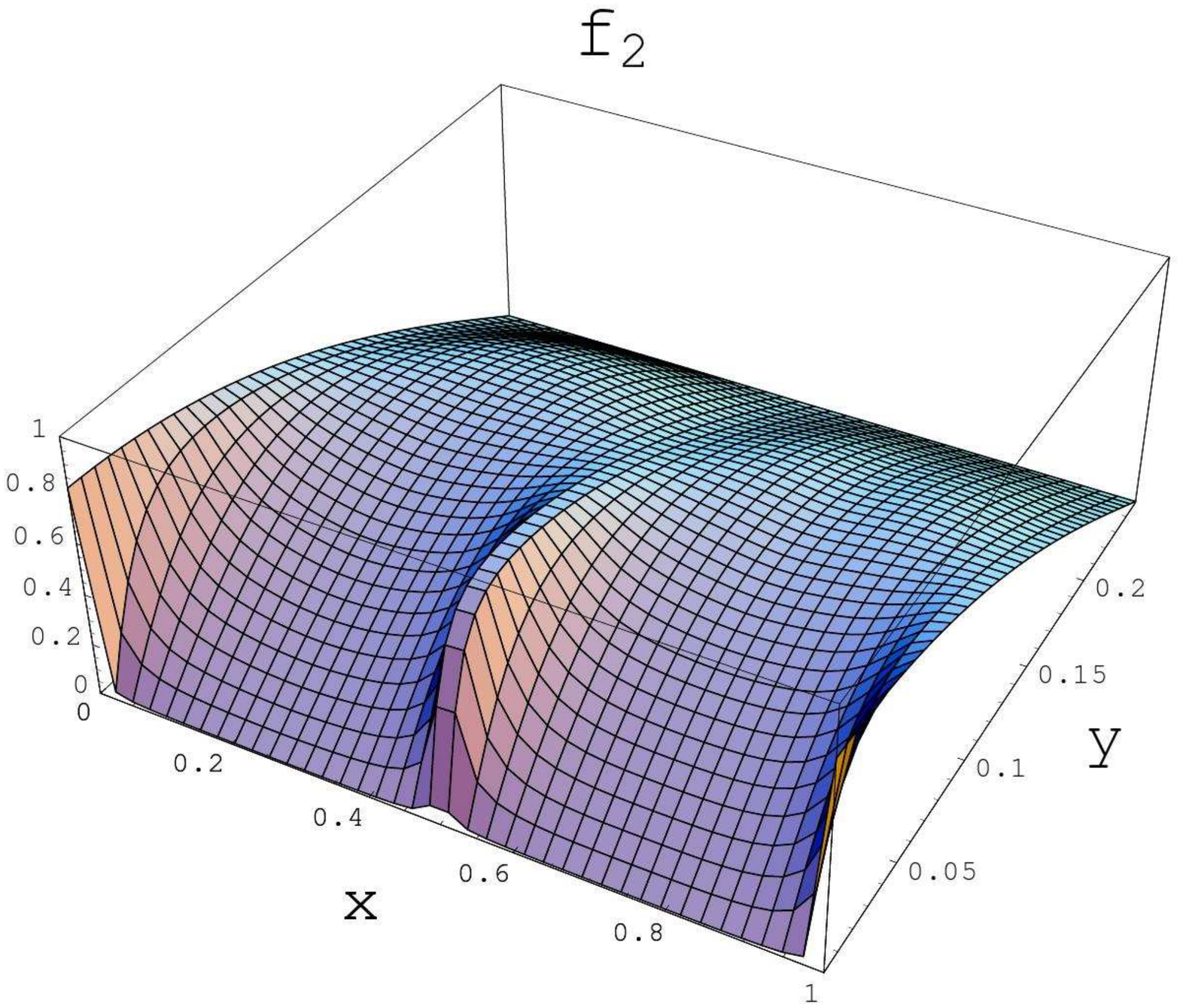}
\caption{Metric factors $f_1$ and $f_2$ for the example of section \ref{example}. }
\label{example1a}
\end{center}
\end{figure}
We can see that the case $N=1,N'=0$, i.e. there is only one asymptotic region,  is impossible as follows: 
$M+M'=2(N+N')$ implies that $M+M'=2$. The first condition from above implies that $M=1,M'=1$. 
However this contradicts the second condition. \\
Hence, in the simplest case, one has at least two asymptotic regions, i.e. $N+N'=2$. 
It follows from (\ref{noofpara}) that $M+M'=4$ and from the second condition above we see that  we 
have $M=2,M'=2$. \\
There are two distinct cases since the boundaries of the 
annulus can be exchanged: in the \emph{planar} case ($N=2,N'=0$),  the two asymptotic regions are on the same boundary 
and in the \emph{non-planar} case ($N=1,N'=1$), the two asymptotic 
regions are on different boundaries.
\begin{figure}[h!]
\begin{center}
\includegraphics[ scale=0.4]{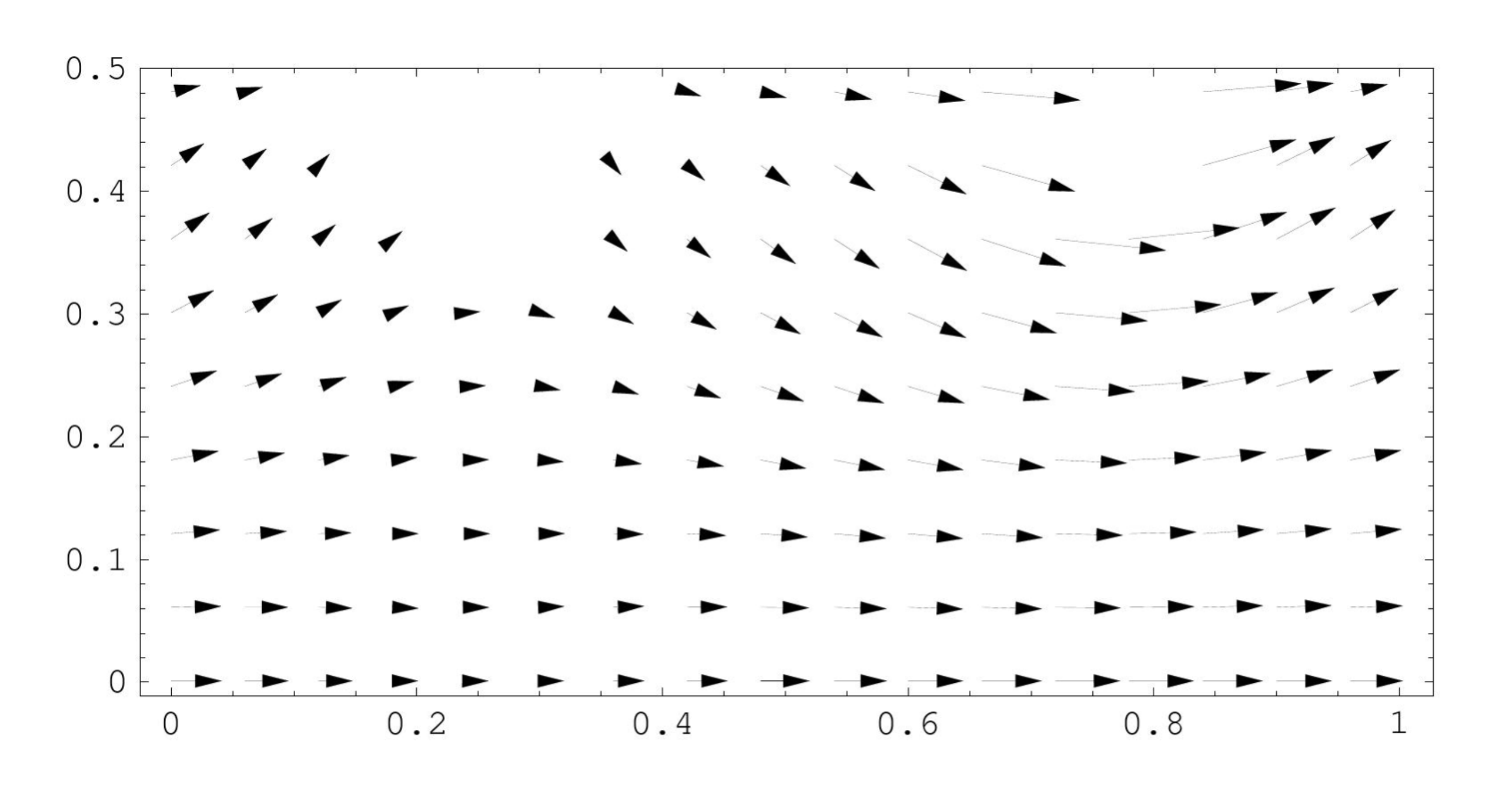}
\caption{Field lines of $f^4_3 \rho \tilde h_a = \partial_a C_K $ for the example of section \ref{example} }
\label{example-D3lines}
\end{center}
\end{figure}
We present plots of some of the metric functions for the planar case and  an annulus with $\tau=i$. 
The poles of $H$  are located on the lower boundary at $x=0$ and $x=1/2$ and have unit residues.
The poles of $A$ on the lower boundary are at $x=1/4,3/4$ and have residues $1$ 
and $2$ respectively. The poles of $A$ on the upper boundary are at $y=1/4,y=3/4$ and have residues $2$ and $1$.
The field lines of the combination $f^4_3 \rho \tilde h_z $ are plotted in figure \ref{example-D3lines}. 

\bigskip

\bigskip

\section{Degeneration and the probe limit}\label{sec4}
\setcounter{equation}{0}
Riemann surfaces with more than one boundary (or with handles) have moduli which correspond to deformations of the surface. 
It is interesting to consider the behavior of our solutions at the boundary of this moduli space, i.e. when the annulus degenerates. 
The degeneration we will consider in this section is given by the shrinking of one of the holes of the annulus to zero size.
 When the inner boundary  shrinks to zero size, it is replaced by a puncture in the interior of the disk. 
Note that each point on the disk can be associated with a particular $AdS_2\times S^2$ slice of the geometry.
 In the parameterization of the annulus defined in (\ref{annua}) this limit corresponds to taking $\tau \rightarrow i \infty$. 
It follows from the formula for the D3-brane charge that in this limit 
$q_{D3}\to 0$ in accord with the expectation that in the probe limit the D3 brane charge becomes negligible in comparison to the charges 
which support the $AdS_3 \times S^3$ geometry. \\
In the degeneration limit, the fundamental domain will be the upper half-plane with identification $w \simeq w+1$.\\
A useful expression in the expansion is given by
\be
{\vartheta'_1(\pi w|\tau) \over \vartheta_1(\pi w|\tau) } = \pi \cot \pi w + 4\pi \sum_{n=1}^\infty{q^{2n} \over 1-q^{2n}} \sin(2\pi n w)
 \label{thetadeg} \ee
In this limit we have
\bea
\lim_{t\to \infty} h_0(w-x ,\bar w-x)  &=& \pi i \big( \cot \pi (w-x)-  \cot \pi (\bar w-x)\Big)  \nonumber\\
\lim_{t\to \infty} h_0(w'+y ,\bar w'+y)  &=& 0
\eea
The function $h_0(w-x,\bar w-x)$ still obeys Dirichlet boundary conditions for real $w$ and has a pole for $w=x$. 
The functions $A$ and $H$ assume the following forms in the degeneration limit,
 \bea H_D &=& \lim_{t \rightarrow \infty} H = \pi i \sum^N_{\alpha=1} r_{H,\alpha} \cot \pi (w-x_{H,\alpha}) + c.c.   \\
 A_D &=& \lim_{t \rightarrow \infty} A = \pi i \sum^M_{\alpha=1} r_{A,\alpha} \cot \pi (w-x_{A,\alpha})    \eea 
Using (\ref{Bform}), the meromorphic function $B$ can be written as,
\bea B_D &=& \lim_{t \rightarrow \infty} B = - { i \pi \over \mathcal{N} } \sum^N_{\alpha=1} {r_{H,\alpha} \over \sin^2 \pi (w-x_{H,\alpha})} \\
 \mathcal{N} &=& e^{i \pi (M-2N)} { \prod^M_{\alpha=1} \sin \pi (w-x_{A,\alpha}) \over \prod^N_{\alpha=1} \sin^2 \pi (w-x_{H,\alpha})  } \eea
It is interesting to note that the contributions to $B$ from terms with $h_0(w'+y ,\bar w'+ y)$ 
will not in general vanish because the residues $r'_{B,\b}$ depend on the modular parameter $\tau$ through the definition
(\ref{Bform}) and may become infinite in the degeneration limit. \\
  
At this point, we intend to analyze our solution near the point to which the upper boundary $\partial \Sigma_2$ has degenerated.
Hence, we need to study
our basic harmonic functions when $\Im w \rightarrow \infty$. Note that this limit is taken only after the degeneration limit $\tau \rightarrow i \infty$. \\ 
It is easy to see that the harmonic functions $H$ and $A + \bar A$ become constants. Moreover, it is possible to show that $\mathcal{N}= \pm 1$ using (\ref{cond-annulus2}), (\ref{cond-annulus3}) and the expansion (\ref{thetadeg}).  
Since $B$ is defined in terms of $\partial_w H$, we get that
\be \lim_{\Im w \rightarrow \infty} B_D = 0  \ee
Hence, the upper boundary has completely disappeared leaving only a zero of $B$. We can map the infinite cylinder to the half plane using the map
\bea
  z= \tan( \pi w)
\eea
Which maps $w=0$  to $z=0$ and $w={1\over 2}$ to $z=\infty$.
The basic harmonic function is expressed in the new coordinates as
\be
\lim_{t\to \infty} h_0(w-x ,\bar w-x)  =  i\pi  {1 + z \hat x \over z- \hat x} + c.c. = 2 \pi {\Im z (1 + \hat x^2) \over |z- \hat x|^2}
\ee
Where $\hat x = \tan x$. In this coordinates, the meromorphic function $B$ is given by
\bea B_D &=& - { i \pi \over \mathcal{N}} \big( z^2 +1 \big) \sum^N_{\alpha=1}r_{H, \alpha}  {1+ \hat{x}_{H,\alpha} \over (z- \hat{x}_{H,\alpha})^2}\\
 \mathcal{N} &=& (1-i z)^{2N-M}  {
\prod^M_{\alpha=1} {  \textstyle z- \hat{x}_{A,\alpha} \over \textstyle \sqrt{ \hat{x}^2_{A,\alpha}+1} }
\over  \prod^N_{\alpha=1} {\textstyle(z- \hat{x}_{H,\alpha})^2 \over \textstyle \hat{x}^2_{H,\alpha}+1 }
} \eea
We can see that $B_D$ has a simple zero in the interior of $\Sigma$ for $z=i$. 
We also note that all the other relevant harmonic functions are strictly positive in the bulk by construction and will have a finite non-zero value at this point.\\
If we expand in polar coordinates on the plane so that $r=|z-i|$, equation (\ref{sol-rho}) gives the following expression for the two-dimensional metric,
\be \rho^2 dz d \bar z \sim {\text{const} \over r^2 }  \big( dr^2 + r^2 d\phi^2  \big) = { \text{const} }    \Big( {dr^2 \over r^2} +  d\phi^2  \Big) \ee
Hence, there is an infinite geodesic distance from the boundary to the point at $z=i$.
The behavior of $\rho$ is the same encountered in \cite{Chiodaroli:2009yw} close to points which are zeros of $B$ but not zeros of $\partial_z H$. 
These points correspond to $AdS_2 \times S^2 \times S^1 \times R$ regions where the $AdS_2$ and $S^2$ spaces have the same radius.  

\bigskip 

\bigskip

\section{General multi-boundary solutions}
\label{sec5}
\setcounter{equation}{0}

The example of the annulus already displayed the new salient properties of solutions 
associated with Riemann surfaces $\Sigma$ with multiple boundary components, 
namely the presence of homology three spheres and the non-vanishing D3-brane 
charges due to the fact  that  $\tilde K$ is not single-valued around the  
non-contractible cycle of the annulus. Furthermore the degeneration of the 
Riemann surface has an interesting   physical interpretation. Note  that many 
of the objects and techniques we employ in this section are also used in 
multi-loop string perturbation theory (see \cite{D'Hoker:1988ta} for a review). 
It is an interesting question whether 
this is a coincidence or points towards a deeper connection of the two subjects.

\subsection{Doubling the Riemann surface}

In  this section we shall generalize the construction to Riemann surfaces with an 
arbitrary number of boundaries but without handles. It is well-known from 
complex analysis \cite{Fay:73} and open string perturbation theory that a Riemann 
surface $\Sigma$ with boundaries can be related to a compact Riemann surface 
$\bar \Sigma$ without boundary which is referred to as the double of $\Sigma$.
Specifically, the double $\bar \Sigma$ must possess an anti-conformal involution,
denoted $\cI$, which is such that $\Sigma = \bar \Sigma / \cI$. For example, 
in the case of the annulus, the double is a rectangular torus, and the involution  
$\cI$ is just complex conjugation.

\sm

We generalize the construction as follows \cite{Fay:73,D'Hoker:1988ta}. 
Let $\Sigma$ be a surface with 
$N_B$ disjoint boundaries, and no handles, or equivalently let $\Sigma$ have the
topology of a sphere with $N_B$ disjoint discs removed. The double $\bar \Sigma$
is a compact Riemann surface of genus $g=N_B-1$ without boundary.
The anti-conformal involution $\cI$ sets $\Sigma=\bar \Sigma/\cI$. 
We choose a canonical homology basis for $H_1(\Sigma)$ of cycles $A_i$ 
and $B_i$ with  $i=1,\cdots, g$ on $\bar \Sigma$ which are adapted to the 
involution $\cI$. Specifically, we identify $N_B-1$ boundary components with 
the A-cycles of $\bar \Sigma$, so that the involution acts on the cycles 
$A_i, B_i$ through,
\bea
\label{ione}
\mathcal{I}(A_i) =A_i \hskip 1in \mathcal{I}(B_i)= - B_i
\eea
Next, we choose a basis of holomorphic differentials $\omega_i$ on $\bar \Sigma$ 
satisfying, 
\bea
\label{itwo}
\mathcal{I}^*(\omega_i (z))= \bar \omega_i (\mathcal{I}(z))
\eea
where $\cI^*$ denotes the pull-back of $\cI$ to differential forms.
Finally, as usual, we choose the differentials $\o_i$ to have canonical 
normalization on the 
$A_i$-cycles, and define the period matrix $\tau_{ij}$ in the standard manner,
\bea
\oint_{A_j}\omega_i =\delta_{ij} \hskip 1in \oint_{B_j} \omega_i =\tau_{ij}
\eea
Conditions (\ref{ione}) and (\ref{itwo}) imply that the period matrix $\tau_{ij}$ 
is purely imaginary. Note that  if the surface $\Sigma$ also had handles, the action 
of the involution $\cI$ on the cycles as well as on the period matrix, would be more 
complicated. We will postpone this case for later work.

\sm

\begin{figure}
\centering
\includegraphics[ scale=0.6]{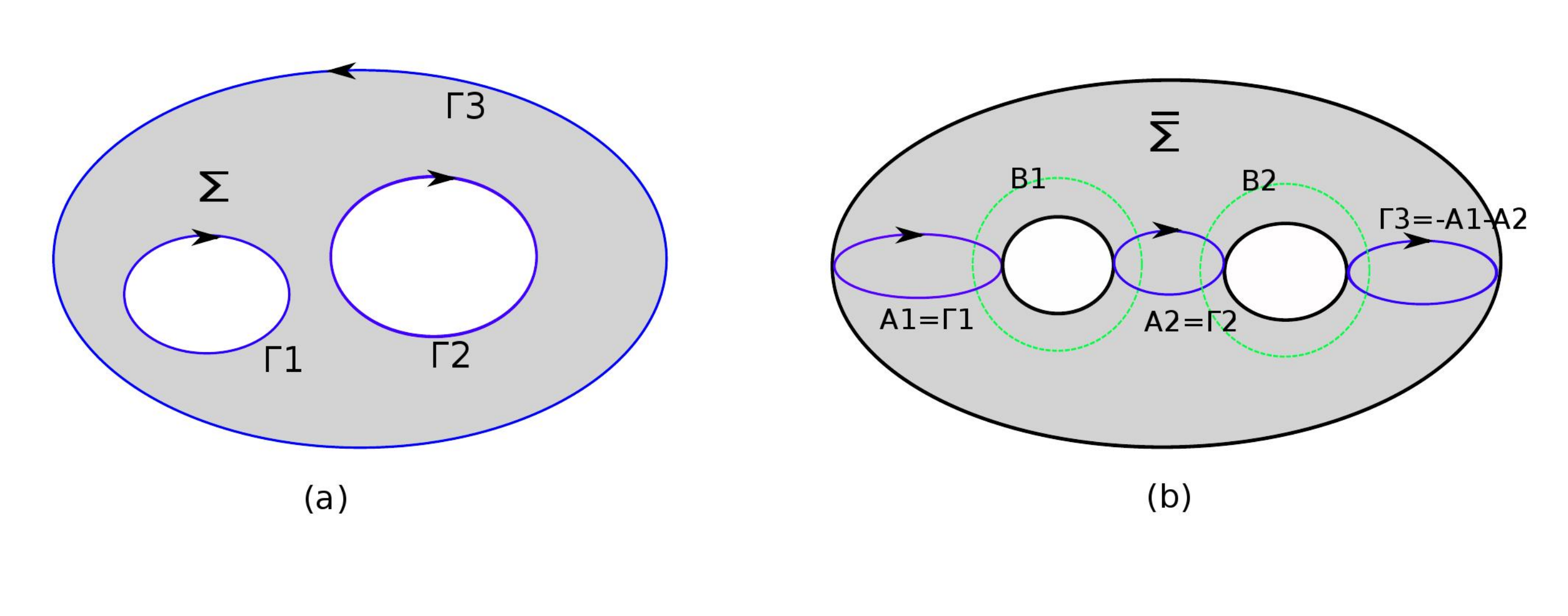}
\caption{(a) A Riemann surface $\Sigma $ with three boundary components, 
(b) and its double $\bar \Sigma$. The boundaries are fixed points under the involution $\cI$}
\label{double1}
\end{figure}

All necessary functions can be constructed from the theta functions  on the double 
$\bar \Sigma$. Higher genus theta functions are defined on a complex torus
${\mathbb{C}}^g / \Lambda $ with period lattice $\Lambda=(I,\tau)$, where
$I$ is the $g\times g $ identity matrix and $\tau$ is a $g \times g$ 
symmetric matrix such that $\Im \tau $ is positive. 
For any $u \in {\mathbb{ C}}^g / \Lambda$, the genus $g$ theta function is  defined by,
\bea 
\Theta(u,\tau) 
= \sum_{N \in {\mathbb{ Z}}^g} \exp 2 \pi i \Big( {1 \over 2} N^t \tau N + N^t u \Big)  
\eea  
Similarly, given $\epsilon, \ \epsilon' \in ({\mathbb{Z} /2})^g$,  the theta functions with 
half-integer characteristic $\a=[\epsilon, \epsilon']^t$, are defined by,
\bea 
\Theta_\alpha (u,\tau) 
= \sum_{N \in {\mathbb{Z} }^g} 
\exp  2 \pi i \left( {1 \over 2} ( N + \epsilon )^t \tau
( N + \epsilon ) + ( N + \epsilon)^t 
( u + \epsilon' )  \right) 
 \label{defthetagen} 
 \eea  
In order to define theta functions on a genus $g$ Riemann surface $\Sigma$, 
one needs to introduce the Abel-Jacobi map $\varphi$, 
\bea 
\varphi: \Sigma \rightarrow J(\Sigma)={\mathbb{C}}^g / \Lambda 
\eea  
which embeds the Riemann surface $\Sigma$ into the Jacobian 
$J(\Sigma) \equiv {\mathbb{C}}^g / \Lambda$. To define the Abel-Jacobi map,
one fixes an arbitrary point $z_0 \in \Sigma$, 
\bea 
\varphi: z  \rightarrow \left( \int^z_{z_0} \omega_1,  \dots , \int^z_{z_0} \omega_g \right) 
\qquad
\text{mod } \Lambda 
\eea 
Thus, $\f(z)$ is a $g$-dimensional complex vector of Abelian integrals.
It is standard to use the notation $\varphi(z-w) \equiv \varphi(z)-\varphi(w)$, 
which defines a quantity that is independent of the choice of $z_0$.
We are now ready to define the theta functions on $\Sigma$,
\bea 
\vartheta(z,\tau) & \equiv & \Theta(\varphi(z),\tau)
\no \\
 \vartheta_\alpha(z,\tau) & \equiv & \Theta_\alpha (\varphi(z),\tau) 
\eea
For each odd half-integer characteristic $\alpha$, there exists a unique
holomorphic $(1/2,0)$ form $h_\a(z)$, which is given by, 
\be 
h_\alpha (z)^2 = \sum^g_{j=1} \partial_j \Theta_\alpha (0,\tau) \, \omega_j(z)   
\ee 
The \emph{prime form} $E(x,y)$ is then  defined as
\be  
E(z,w) = {\vartheta_\alpha (z-w,\tau) \over h_\alpha(z) h_\alpha(w)} \label{defE} 
\ee 
The prime form is independent of the choice of odd characteristic $\alpha$;  it is 
antisymmetric $E(z,w)=-E(w,z)$; and vanishes only on the diagonal $w=z$ .  
In a local coordinate system near the zero at $w=z$, the prime form behaves as  
\be
\lim_{w\to z} {E(z,w) \sqrt{dz \, dw} \over z-w } = 1 + \cO \left ( (z-w)^2 \right ) 
\ee
i.e. $w=z$ is a simple zero. For fixed $z$, $E(z,w)$ defines a multi-valued 
holomorphic differential form of weight $-1/2$. The prime form is single-valued 
around $A_j$-cycles, and has the following monodromy around $B_j$ cycles,
\be 
E(z+B_j,w) = - E(z,w) \exp \left \{ - i \pi \tau _{jj} - 2 \pi i \f _j (z-w) \right \}
\ee 
A further differential form $\sigma (z)$  will be needed to construct solutions.  
It is defined by,
\be 
\sigma(z) = \exp \Big\{  - \sum^{g}_{j=1} \int_{A_j} \omega_j(y) \log E(y,z)  \Big\} 
\label{def-sig} \ee
The differential  $\sigma(z)$ has weight $(g/2,0)$, and has neither poles 
nor zeros on  $\bar \Sigma$. It is single-valued around $A_j$-cycles, and 
has the following monodromy around  $B_j$ cycles\footnote{The overall sign  in the monodromy is immaterial  for our purpose, as only $\sigma(z)^2$ appears in the expressions  used in the following section.}:
\be 
\sigma (z+B_j)= \pm \sigma (z) \exp 2 \pi i \left ( {g-1 \over 2} \tau_{jj} - \Delta_j  
+ (g-1) \f_j ( z  ) \right ) 
\ee 
where 
\bea
\Delta_j= \half - \half \tau_{jj} + \sum^{g}_{k\not = j } \int_{A_k}  \omega_k \varphi_j  
\eea
is the vector of Riemann constants. Note that $\Delta _k$ depends upon the 
reference point $z_0$.

\subsection{Construction of the basic harmonic function}

The key tool in the construction of solution for the case of the annulus was the 
existence of a basic positive harmonic  function $h_0$, whose associated meromorphic
part is holomorphic in the interior of $\Sigma$, and has a single simple pole on the 
boundary of $\Sigma$. All meromorphic and harmonic functions needed for the 
annulus were then constructed as linear combinations of this basic function.

\sm

This method may be generalized to the case of multiple boundary components,
whose double is a surface $\bar \Sigma$ of higher genus. Additional care will
be needed for higher genus in taking proper account of the weight of  
differential forms (while for the annulus, all meromorphic differentials could be 
canonically identified with meromorphic functions). To construct a harmonic 
function on a genus $g$ surface, we need a meromorphic function, i.e. a form of 
weight 0, which has simple poles only. 
Such objects arise in Riemann surface theory as Abelian integrals of the second kind,
obtained as indefinite line integrals of  Abelian differentials of the second kind, 
with double poles only. The Abelian differential of the second kind with a double 
pole at the point $p$ (but holomorphic elsewhere), and vanishing $A_i$-cycles,  
is unique, and is given in terms of the prime form by,
\bea
\omega (z,p) = \p_z \p_p \ln E(z,p) \hskip 1in \int _{A_i} dz \, \o(z,p) =0
\eea
It is single-valued on $\bar \Sigma$. The corresponding Abelian integral is given by,
\bea
 \int ^z _q dy\,  \o (y,p) = \p_p \ln \left ( { E(z,p) \over E(q,p) } \right )
\eea
For our purposes, however, the requirement on $\o(z,p)$ 
of vanishing integral around $A$-cycles is too restrictive. Relaxing this condition
allows us to add to $\o(z,p)$ any linear combination of the holomorphic Abelian 
differentials $\o_i$, with $i=1,\cdots, g$. The corresponding Abelian integral
is then given by,
\bea
Z  (z|p,q) = \p_p \ln \left ( { E(z,p) \over E(q,p) } \right ) 
+ \sum _{i=1} ^g C_i (p,q) \int ^z _ q \o_i
\eea
for an as yet undetermined $C_i(p,q)$. Clearly, $Z(z|p,q)$ still has a simple pole 
at $z=p$, and a simple zero at $z=q$, but now has non-trivial monodromy around
both $A_i$- and $B_i$-cycles, 
\bea
Z(z+A_j|p,q) & = & Z (z|p,q) + C_j (p,q)
\no \\
Z(z+B_j|p,q) & = & Z (z|p,q) + 2 \pi i \o_j (p) + \sum _{i=1}^g C_i (p,q) \tau _{ij}
\eea
Some care is needed in specifying the derivatives with respect to the locations $p$
of the poles. Indeed, the poles of the harmonic functions $A\pm \bar A, 
B \pm \bar B, H$ and $K$ need  to be on the boundary of $\Sigma$, 
namely at points $p$ which are invariant under the involution $\cI(p)=p$. 
Therefore, the derivatives with respect to $p$ which enter into the definition 
of $Z(z|p,q)$ must be taken along the boundary only. In a suitable local 
coordinate system, $(w, \bar w)$, the boundary $\p \Sigma$ may be described 
locally by $\Im (w)=0$. In this coordinate system, the differentiation in $p$ along 
the boundary is defined as the  derivative with respect to $\Re(w)$. 
All derivatives and differentials in $p$ will be understood in this manner. 

\subsubsection{Satisfying vanishing Dirichlet boundary conditions}

We are now ready to define the basic harmonic functions $h(z|p)$ on $\Sigma$, 
\bea
h (z|p) \equiv  i\Big (  Z (z|p,q)- Z(\mathcal{I}(z)|p,q) \Big )
\eea
where the points $p,q$ are on the boundary of $\Sigma$ and satisfy 
$\cI(p)=p, ~ \cI(q)=q$. We have suppressed the dependence on the
point $q$ in $h(z|p)$, because we shall show shortly that, even though $Z(z|p,q)$ 
depends upon $q$, this dependence cancels out of $h(z|p)$.
By construction, the harmonic function $h(z|p)$ is real, and has a single simple 
pole at $z=p$. 
It remains to determine the $C_i(p,q)$ so that $h(z|p)$ consistently obeys vanishing 
Dirichlet boundary conditions on all $g+1=N_B$ boundary components of $\Sigma$. 
If $z,p,q$ are all on the same boundary component of $\Sigma$, it is 
manifest that $h(z|p)=0$. We need to ensure that $h(z|p)=0$
continues to hold when $z$ is on a boundary component  different from that of $p$.
Points on different boundary components may be mapped into one another 
by moving the points through $\Sigma$ along sums of half-$B$-periods.

\sm

Recall that $\p \Sigma$ has $N_B=g+1$ disconnected components $\G_i$, 
and that we have identified $\G_i = A_i$ for $i=1,\cdots, g$. 
Concretely, let $z,p,q\in \G_k$, and let $z' \in \G_{k'}$. Comparing
the values of $h(z|p)$ on these different boundary components gives,
\bea
\label{hpq1}
h (z'|p) - h(z|p) = i \left ( \int _z ^{z'} -  \int ^{\cI(z')} _{\cI(z)} \right ) 
\left ( dy \, \o(y,q) +  \sum _{i=1}^g C_i(p,q) \o_i \right )
\eea
It suffices to consider nearest neighbor cycles, with $k'=k+1$, for 
$k =1, \cdots, g$, all others being given by linear combinations of these.
The difference of the integration contours entering (\ref{hpq1}) 
then precisely coincides with the cycle $B_k$, and the integrals may
all be carried out to give, 
\bea
h (z'|p) - h(z|p) = - 2 \pi \o_k(p) +i \sum _{j=1}^g C_j (p,q) \tau_{jk}
\eea
The harmonic function $h(z|p)$ will consistently obey vanishing 
Dirichlet boundary conditions on all boundary components provided 
$h (z'|p) - h(z|p) =0$ for any pair $z,z'$ on the boundary of $\Sigma$. This gives $g$ conditions,
which determine $C_i(p,q) $ uniquely,
\bea
C_i(p,q) = - 2 \pi i \sum _j \left ( \tau ^{-1}\right )_{ij} \o_j(p)
\eea
Note that $C_i(p,q)$ is indeed independent of $q$, and is a well-defined holomorphic
Abelian in $p$. Henceforth, we shall also suppress the $q$-dependence of $C_i$
and simply refer to it as $C_i(p)$. Since $\tau_{ij}$ is purely imaginary, the 
quantities $C_i(p)$ are real. Since $C_i(p)$ is real when $p,q$ are on the boundary
of $\Sigma$, it follows that all $q$-dependence cancels out of $h(z|p)$, as promised.

\sm

Although $Z (z|p,q)$ has non-trivial monodromy 
$C_k(p)$ around any cycle $A_k$, this monodromy is clearly real, and 
cancels out for $h(z|p)$, which is thus single-valued around 
every $A_k$-cycle.  The function $\tilde h (z|p)$ is the harmonic
dual to $h(z|p)$, and is given by
\bea
\tilde h(z|p) =    Z (z|p,q)+ Z(\mathcal{I}(z)|p,q) 
\eea
Its monodromy around an $A_i$-cycle is given by $2C_i(p)$, and generally
is non-vanishing. The harmonic function $\tilde h(z|p)$ is defined only up
to an additive $z$-independent function, which we choose so as to
cancel the $q$-dependence of $\tilde h(z|p)$, as the notation indeed indicates.
Note that $h(z|p)$ transforms as a one-form with respect to $p$.
This concludes our construction of the basic harmonic functions  $h(z|p)$,
and $\tilde h(z|p)$.

\subsection{Construction of  the supergravity solution}

The supergravity solutions are expressed in terms of the harmonic
functions $A\pm \bar A, B \pm \bar B, H$, and $K$. In this subsection,
we shall now determine these functions for a surface $\Sigma$ with 
an arbitrary number $N_B=g+1$ of boundary components, but no handles.

\subsubsection{The harmonic function $H$}

The function $H$ must be single-valued and positive on $\Sigma$, and obey 
vanishing Dirichlet boundary conditions.  We take $H$ to have $N_i$ poles 
$p_{H,ij}$ on boundary component $i$, labeled by $j=1,\cdots, N_i$.  
The corresponding residues $r_{H,ij}$ must be positive, but are 
otherwise left undetermined,
\bea 
\label{Hform1}
H = \sum^{N_B}_{i=1} \sum _{j=1}^{N_i} r_{H,ij} h(w|p_{H,ij})
\eea
Note that each pole $p_{H,ij}$ of $H$ corresponds to an asymptotic $AdS_3 \times S^3$ region.
 Hence, the full supergravity solution will have a total of $\sum _{i=1}^{N_B} N_i$ 
asymptotic  $AdS_3 \times S^3$ regions.

\subsubsection{The harmonic functions $A\pm \bar A$}

The harmonic function $A + \bar A$ must be single-valued and 
positive in $\Sigma$, and obey vanishing Dirichlet boundary conditions. 
The conjugate function $-i(A - \bar A)$ must also be single-valued on $\Sigma$.  
As a result, these functions take on the following form,
\bea 
\label{Aform1}
(A + \bar A) 
&=& \sum^{N_B}_{i=1} \sum _{j=1}^{M_i} r_{A,ij} h(w|p_{A,ij})
\no \\
-i(A - \bar A) 
&=& \sum^{N_B}_{i=1} \sum _{j=1}^{M_i} r_{A,ij} \tilde h(w|p_{A,ij})
\eea 
Single-valuedness of $-i(A-\bar A)$ around each $A_k$-cycle  imposes a relation 
between the residues.
Using (\ref{mono1}), one obtains the monodromy of this function,
\bea
\sum^{N_B}_{i=1} \sum _{j=1}^{M_i} r_{A,ij} C_k(p_{A,ij})
\eea
Hence, $-i(A-\bar A)$ will be single-valued provided the following relation is obeyed,
\bea\label{realapab}
\sum^{N_B}_{i=1} \sum _{j=1}^{M_i} r_{A,ij} \o_k(p_{A,ij})= 0
\eea
As stated before, the basic harmonic function $h(z|p)$ transforms as a one-form in $p$. As a result, the residues $r_{H,ij}$ and $r_{A,ij}$ are $-1$-forms 
in $p_{H,ij}$ and $p_{A,ij}$ respectively.

\subsubsection{The harmonic functions $B\pm \bar B$}

$\bullet$ The meromorphic function $B$ must have the same poles as $A$
(according to the regularity condition {\bf R1}), and the same zeros as $\partial_w H$ 
(according to {\bf R4}). Moreover, both $B+\bar B$ and $-i(B- \bar B)$ must be single-valued. Thus, these functions can be expressed as
\bea 
\label{Bform1}
(B + \bar B)
&=& \sum^{N_B}_{i=1} \sum _{j=1}^{M_i} r_{B,ij} h(w|p_{A,ij})
\no \\
-i(B - \bar B)
&=& \sum^{N_B}_{i=1} \sum _{j=1}^{M_i} r_{B,ij} \tilde h(w|p_{A,ij})
\eea 
The functions $B+\bar B$ and $-i(B-\bar B)$ are not subject, however, to any positivity 
condition, and are allowed to change sign inside $\Sigma$. We get the following 
expression for $B$:
\bea 
B(w) =  { \partial_w H (w) \over \mathcal{N}(w)}
\eea
Here, $\cN(w)$ is a single-valued meromorphic form of weight $(1,0)$, 
whose zeros and poles are determined by the requirement that $B$ and 
$\p_wH$ have common zeros and $B$ and $A$ have common poles. 
Thus, $\cN$ must cancel the (double) poles of $\p_wH$, and reproduce 
the poles of $A$. Before constructing $\cN$, we
shall first derive the final harmonic function $K$.

\subsubsection{The harmonic function $K$}

The harmonic function $K$ has the same poles as $A+ \bar A$ and the 
residues are given by the condition {\bf R1}.
The expression is fixed to be:
\bea
\label{Kform1}
K = \sum^{N_B}_{i=1} \sum _{j=1}^{M_i} ~ {r_{B,ij}^2 \over r_{A,ij}} ~ h(w|p_{A,ij})
\eea
In this expression, $K$ is manifestly positive everywhere on $\Sigma$.
Note that the dual harmonic function $\tilde K$ does not need to be single-valued. 

\subsubsection{Construction of $\cN$}

The meromorphic form  $\mathcal{N}(z)$ must satisfy the following properties:
\begin{enumerate}
 \item $\mathcal{N}(z)$ has the same (double) poles as  $\partial_w H$, and has 
  zeros where $A$ has poles.
\item $\mathcal{N}(z)$ is real on all boundaries.  
\item $\mathcal{N}(z)$ is a $(1,0)$-form so that $B$ transforms as a scalar.
\item $\mathcal{N}(z)$ is single-valued around $A$-cycles.
\end{enumerate}
To carry out the explicit construction of $\cN(w)$, it is convenient to work with 
meromorphic differential one-forms for which all $2g-2$ zeros are at prescribed points 
on $\Sigma$. Generally, such one-forms will have monodromy around both $A$-
and $B$-cycles, but it will suffice here to restrict to forms which are single-valued
around $A$-cycles. These may be conveniently constructed out of the 
prime form $E(z,w)$ and the holomorphic form $\sigma (z)$ of weight $g/2$ 
introduced in (\ref{def-sig}). Any holomorphic one-form $\nu(z)$ with zeros at points $p_\a$, 
with $\a = 1, \cdots , 2g-2$, and vanishing $A$-periods may be expressed as follows,
\bea
\nu (z) = \sigma (z)^2 \prod _{\a=1}^{2g-2} E(z,p_\a)
\eea
The monodromy around any cycle $A_k$ vanishes, while around a 
cycle $B_k$ it is given by,
\bea
\nu (z +B_k) = \nu (z) 
\exp 2 \pi i \Big \{   \f _k (p_1 + \cdots + p_{2g-2})  - 2 \Delta_k  \Big \}
\eea
The condition for the vanishing of this monodromy precisely coincides with 
the requirement that $p_1 + \cdots + p_{2g-2}$ is the canonical divisor,
as should have been expected.

\sm

We take the following Ansatz: 
\be 
\mathcal{N}(z)= \exp \left ( {\pi i } \theta ^t \varphi (z) \right ) 
{\prod^{N_B}_{i=1} \prod^{M_i}_{j=1} E(z,p_{A,ij})\over \prod^{N_B}_{i=1} \prod^{N_i}_{j=1} E(z,p_{H,ij})^2  }  {  \sigma(z)^2 }
\ee
The poles of $A$ and $H$, respectively given by $p_{A,ij}$ and $p_{H,ij}$, 
are labeled by an index $i$ which refers to the boundary, and a label $j$ which 
refers to the point on boundary $i$. 
The functions $A$ and $H$ have respectively $M_j$ and $N_j$ poles on  
boundary $j$. It is immediate to see that $\cN$ has the desired poles and zeros,
and only those.  Furthermore,  $\theta$ is a $g$-vector of constants. Recall that 
$E(z,w)$ and $\sigma (z)$ are, and that $\cN$ must be single-valued around 
$A$-cycles. This puts a strong restriction on the components of the vector 
$\theta$,
\bea
\theta _k \equiv 0 \qquad ({\rm mod} ~2)
\label{condAcycle}\eea  
Next, $\cN(z)$ must be a form of weight $(1,0)$ in $z$; this requires
the following relation between the number of poles of $\p_wH$ and the 
number of zeros of $A$,
\bea\label{conmultgen}
2g-2 =   \sum _{i=1}^{N_B} (M_i - 2 N_i)  
\eea
It remains to ensure  that $\cN(z)$ is real on all boundary components. 

\subsubsection{Reality of $\cN$ on all boundary components}

We start by noting that $E(z,w)$ is real when both points $z,w$ are on the 
same boundary.  Next, we will evaluate $E(z,w)$ when $z,w$ are on different boundary components. 
This may be done using the following identity:
\be 
\Theta \Big[ \begin{array}{c} \epsilon \\ \epsilon'  \end{array} \Big] (u + \tau \delta,\tau) 
= \exp \Big[- \pi i \Big( \delta^t  \tau \delta  + 2 \delta^t  \epsilon'  
+ 2 \delta^t u \Big)  \Big] 
\Theta \Big[ \begin{array}{c} \epsilon+\delta  \\ \epsilon'  \end{array} \Big] (u, \tau) 
\label{thetaid}
\ee
with $u \in \mathbb{C}^g/\Lambda$, and $\delta \in (\mathbf{Z}/2)^g$. 
This identity is the analog of the half-period relation (\ref{halfperiod1}-\ref{halfperiod2}).
The integral of a holomorphic differential along an half $B$-cycle may
be expressed in the following from,
\be 
\int_{B^+_j} \omega_i = R_{ij} + {1 \over 2} \tau_{ij} 
\label{eq-halfperiod}
\ee
where $R$ is a real matrix and $B^+_j$ is the upper half of the $B_j$ cycle 
so that $B_j=B^+_j - \cI(B^+_j)$. We can then use (\ref{thetaid}) and 
(\ref{eq-halfperiod}) to find the multiplicative factor acquired by $E(z,p_{H,jk})$ 
when $z$ is taken on boundary component $i$, and $p_{H,jk}$ is on  boundary
component $j$. We construct a curve $C_{ij}$ going from the i-th boundary to the j-th boundary as,
\be 
C_{ij}= \sum_k c_{ij,k} B^+_{k} = c_{ij}^t B^+
\ee
Here we have introduced a tensor $c_{ij,k}$, with $i,j=1 \ldots N_B=g+1$ and $k=1 \ldots g$. 
$C_{ij}$ is constructed as a linear combination of half B-cycles and $c_{ij,k}$ are the expansion coefficients. 
More explicitly, the $c_{ij,k}$ are given by,
\be 
c_{ij,k} = \left\{ \begin{array}{cl} 1 & i\leq k<j \\ -1 & j \leq k<i \\ 0 & \text{otherwise}   \end{array} \right.  
\ee 
It is easy to see that $C_{i+1 \; j}= C_{ij}- B^+_i$. Hence, the tensor $c$ must obey the identity, 
\be c_{i+1 \; j ,k} = c_{ij,k}-\delta_{ik}  \label{iddelta}
\ee 
where $\delta_{ij}$ is the Kronecker delta. With this notation we have that
\be E(x,p_{H,jk}) = \exp \Big[\pi i \Big( {c_{ij}^t  \tau c_{ij} \over 4 } + {c_{ij}^t \epsilon'  }
+ c_{ij}^t \varphi(x- p_{H,jk}) \Big)  \Big] { \Theta \Big[ \begin{array}{c} \epsilon - {c_{ij} \over 2} \\ \epsilon'  \end{array} \Big] 
(\int^{x}_{p'_{H,jk}}  \omega - R c_{ij}, \tau ) \over h \Big[ \begin{array}{c} \epsilon \\ \epsilon'  \end{array} \Big] (x)
 \; h \Big[ \begin{array}{c} \epsilon \\ \epsilon'  \end{array} \Big] (p_{H,jk})  } \label{Ephase} \ee
where $p_{H,jk}$ and $p'_{H,jk}$ are the endpoints of $C_{ij}$ so that $p'_{H,jk}$ is on the same boundary of $x$. 
The argument of the theta function is manifestly real since it can be evaluated using a path which stays on the i-th boundary.
From the definition (\ref{defthetagen}) we see that
 the theta function is real as long as the argument is real. 
Moreover, the one-form $h_\alpha (x)^2$ is real when  $x$ is on one of the boundaries.
The phase of $h_\alpha (x)$ then only depends on the characteristic $\alpha=[\epsilon, \epsilon']^t$ and must have the
effect to cancel any dependence from $\alpha$ in the exponential of (\ref{Ephase}).  
Therefore, the phase of $E(x,p_{H,jk})$ is given by the exponential
\be \exp \Big\{  \pi i  c_{ij}^t \varphi(x-p_{H,jk}) \Big\} \ee 
A similar analysis can be repeated for $\sigma(x)^{2}$. Its phase is given by the exponential
\be \exp \Big\{  2 \pi i \sum^{g}_{j=1} \sum^{g}_{k=1} c_{ij,k}  \int_{A_j} \omega_j (y) \varphi_k(y-x) \Big\} \ee
If $x$ is on the $i$-th boundary, we get that 
the contribution to the phase of $\mathcal{N}(x)$  coming from the prime forms is given by the following expression,
\be - 2 \pi \sum^{N_B}_j \sum^{N_j}_k c_{ij}^t \Re \varphi(x- p_{H,jk})  
+  \pi \sum^{N_B}_j \sum^{M_j}_k c_{ij}^t  \Re \varphi(x- p_{A,jk}) + {\pi } \theta^t  \Re \varphi(x)  \ee
while the contribution coming from $\sigma(x)^2$ is the following,
\be 2 \pi \sum^{g}_{j=1} \sum^{g}_{k \neq j} c_{ij,k}  \Re \Big(  \int_{A_j} \varphi_k(y)  \omega_j(y) - \varphi_k(x)   \Big)  \ee
In conclusion, to have a real $\mathcal{N}(x)$ on all boundaries, the position-dependent part of the phase must be identically zero, 
\be 2 \Big( \sum^{N_B}_{j=1}  (M_j-2N_j) c^t_{ij} +  \theta^t  \Big) \Re \varphi(x) - 4 \sum^{g}_{j=1} \sum^{g}_{k \neq j} c_{ij,k} \Re \varphi_k(x) 
 = 0 \quad (\text{mod } 2) \label{gencond1} 
\ee
Moreover, the constant part of the phase must add up to an integer multiple of $\pi$,  
\be
 4 \sum^{N_B}_{j=1} \sum^{N_j}_{k=1} c_{ij}^t  \Re \varphi(p_{H,jk})- 2 \sum^{N_B}_{j=1} \sum^{M_j}_{k=1} c_{ij}^t \Re \varphi(p_{A,jk}) 
 + 4 \sum^{g}_{j=1} \sum^{g}_{k \neq j} c_{ij,k} \Re \int_{A_j} \varphi_k \omega_j  = 0 \quad (\text{mod } 2) \qquad 
 \label{gencond2} \ee
Furthermore, we note that if $j \neq 1$ then,
\be c_{1j,k} = \sum^{j-1}_{l=1} \delta_{lk} \label{iddelta2}\ee
Evaluating equation (\ref{gencond1}) for $i=1$ and using the identity (\ref{iddelta2}), as well as the fact that the $\varphi_k(x)$ are linearly independent, one  can solve for $\theta_k$,
\be \theta_k =  \sum^{N_B}_{j=k+1} \big( 2 N_j -  M_j \big) + 2 g- 2k   \ee 
We note that (\ref{condAcycle}) can be satisfied only if all the $M_j$ are even.\\
Moreover, we subtract equation (\ref{gencond1}) evaluated for $i$ and $i+1$  and use the identity (\ref{iddelta}) to obtain
\be \sum^{N_B}_j M_j = 2 \sum^{N_B}_j N_j + 2 g -2 \ee 
which is exactly (\ref{conmultgen}). The identity (\ref{iddelta2}) allows us to rewrite the first condition 
from (\ref{gencond2}) as follows,
\be  
\sum^{N_B}_{j=2} \sum^{j-1}_{l=1} \left\{ 4 \sum^{N_j}_{k=1} \Re \varphi_l (p_{H,jk}) - 2 \sum^{M_j}_{k=1} \Re \varphi_l (p_{A,jk}) \right\}
+ 4 \sum^{g}_{j=2} \sum^{j-1}_{l=1}  \Re \int_{A_j} \varphi_l \omega_j = 0 \quad (\text{mod } 2) \qquad  
\ee
We can also use (\ref{iddelta}) to reduce the remaining $g$ conditions from (\ref{gencond2}) to
\be  \label{realnfun}
  4 \sum^{N_B}_{j=1} \sum^{N_j}_{k=1} \Re \varphi(p_{H,jk}) - 2 \sum^{N_B}_{j=1} \sum^{M_j}_{k=1} \Re \varphi(p_{A,jk}) + 4 \Re \Delta =0  \quad (\text{mod } 2)
   \ee
where $\Delta$ is the vector of Riemann constants. These conditions reduce to (\ref{cond-annulus1}), (\ref{cond-annulus2}) and (\ref{cond-annulus3}) in case of the annulus.

\subsection{Regularity and properties of the solution}

The solutions introduced in this sections obey to the regularity conditions R1-R4 by construction. 
If we re-define,
\be h_{\alpha}= h(w | p_{A,ij}), \qquad \alpha = 1,2, \ldots, \sum_{j} M_j \ee
it is possible to repeat the argument of section \ref{secR5} to reduce the regularity inequality R5 to Schwartz's inequality. 
As seen before, the inequality will hold strictly provided that $A$ and $B$ are not proportional, 
but  $B$ cannot be proportional to $A$ because it admits zeros in the interior of $\Sigma$ while $A$ is strictly positive in $\Sigma$ by construction.
Counting the number of independent parameters  of the solution of genus $g$ is instructive. 
The harmonic function $H$ contains $2\sum_i N_j$ parameters. The harmonic function $A+\bar A$ contains $2\sum_j M_j+1$ parameters.
 Using (\ref{conmultgen}) to replace $\sum_j M_j$ gives  $6\sum_ i N_j + 4g -3$ parameters which are associated with the harmonic functions. 
 The condition  that $A-\bar A$  is single-valued  around the $g$ A-cycles (\ref{realapab}),  imposes $g$  constraints.
 Furthermore, the reality conditions on ${\cal N}$  (\ref{realnfun}) impose an additional $g+1$ conditions. The constant $c$ and the real additive constant in $\tilde K$ provide two additional parameters. Lastly, there are $g(g+1)/2$ real moduli of the Riemann surface.  Putting this together we get
\bea
{\rm no. \; of \; parameters:} \; &=&  6\sum_ i N_j+{g(g+1)\over 2} + 2g-2
\eea
We can naturally attribute six parameters to each asymptotic region: F1, D1, NS5 and D5 brane charges and the value of two massless scalars 
in the asymptotic region. The other parameters are related to the moduli of the Riemann surface
 and the presence of non-contractible cycles (supporting D3-brane charge) and homological three-spheres 
(supporting five-brane charge). Note that the number of parameters grows quadratically with the genus of the Riemann 
surface leading to the possibility of a \emph{bubbling Janus} solution with many boundaries. 

\bigskip
\bigskip

\section{Discussion}\label{sec6}
\setcounter{equation}{0}

In this paper we expanded the class of half-BPS solutions of type IIB string theory that are  locally asymptotic 
to $AdS_3\times S^3\times M_4$ and  were constructed in an earlier paper \cite{Chiodaroli:2009yw}. 

\sm

We constructed solutions in which the Riemann surface $\Sigma$ has an arbitrary number of
boundaries, and studied the simplest case of the annulus in detail. In this case, the double surface is a (square) torus,
all harmonic functions can be constructed in terms of Jacobi theta functions and the regularity conditions can be solved explicitly.
The requirement that the meromorphic function  $B$ obeys Dirichlet boundary conditions on both boundaries imposes additional constraints 
on the parameters of the solution.  
The main new physical feature of the solutions is the presence of  non-vanishing three-brane charge.
This fact is related to the existence  of  a  non-contractible cycle on the annulus. 
Moreover, there is an additional three-sphere in the geometry which is  not associated with any asymptotic region. 

\sm

We studied the degeneration limit where the inner boundary shrinks to zero size and disappears leaving the disk with a puncture.
The geometry near this puncture is given by an $AdS_2 \times S^2 \times S^1 \times R$ infinite throat. 
Note that we did not scale any of the other parameters of the solution in the degeneration limit.
The existence of other scaling limits with different asymptotics is an interesting open question. 

\sm

The solution for  the annulus was generalized  to Riemann surfaces with  an arbitrary number of holes. 
The construction involves   well known functions on higher genus Riemann surfaces, 
such as  theta functions and   prime forms.  
We constructed the harmonic functions and solved the constraints imposed by reality and regularity. 
In the multi-boundary case, there are many non-contractible cycles coming from the $A$-cycles of the 
double Riemann surface, and each of them can be associated with a non-vanishing D3-brane charge. \\

\noindent There are many interesting features of our solutions and possible extensions that deserve further study:

\begin{itemize}
\item The supergravity solutions we have found are dual to defect and interface theories in two-dimensional conformal 
field theories. Even the simplest case in which $\Sigma$ can be mapped into the disk or upper half-plane has interesting 
solutions. A possible direction for further research is to investigate the holographic duals for the solutions we have found.
In particular, it would be interesting to find out whether the fact that $H$ can have poles 
(corresponding to asymptotic AdS regions) on different boundaries have an interpretation in the dual CFT.

\item Note that the expression for axion (\ref{axionform}) in terms of the harmonic functions
\bea
\chi &=& {i\over 2} \Big( (A - \bar A) +{\bar B^2 -B^2 \over K} \Big)
\eea
is very similar to the formula for $C_K$ (\ref{fourfcharge})  with $K$ replaced by $A + \bar A$. Hence, it seems possible to 
introduce seven-brane charge by dropping the requirement that the harmonic function $i(A-\bar A)$ is single-valued.
 Note that, in this case, the three-form anti-symmetric tensor fields also have non-trivial monodromies. For example, if one chooses the parameters of the harmonic 
function $A$ on the annulus  such that $\chi\to \chi+1$ as $w\to w+1$, then one finds that $F_{3}\to  F_{3}+ H_{3}$.  
Where $F_3$ and $H_3$ are the R-R and NS-NS three-form anti-symmetric tensor fields respectively. 
These are exactly the monodromies one would expect for a single D7-brane \cite{Greene:1989ya,Gibbons:1995vg}.

\item It is interesting that the machinery of higher loop (open and closed) string perturbation 
theory was very useful in constructing the holographic duals of interfaces and defects in two-dimensional conformal 
field theories. This might be a coincidence dictated by the fact that in both cases Riemann surfaces are a basic ingredient.
 However,  it is not inconceivable that there is a deeper connection, possibly indicating a relation between holographic defects and
 open strings in the effective string description of the D1/D5 system \cite{Das:1996jy}

\item In this paper we considered Riemann surfaces without handles. 
In principle, the methods employed in section \ref{sec5} can be used also for Riemann surfaces $\Sigma$ with $n$ boundaries and 
$h$ handles, where the double $\bar \Sigma$ will be a compact Riemann surface of genus $n-1+ 2h$. In this case the period matrix
 is not purely imaginary, and there are additional constraints from the requirement that our harmonic functions are single-valued.  
We leave the investigation of this case, and the very interesting question of its physical significance, 
for future work.
\item In section \ref{sec4} we discussed the degeneration of the annulus where the inner boundary shrinks to zero size 
and we found that in this limit an extra $AdS_2\times S^2 \times S^1 \times R $ asymptotic region appears. 
 We expect that the same interpretation will hold for surfaces with many boundaries where a boundary closes off. 
It is well known from string perturbation theory (see e.g. \cite{D'Hoker:1988ta}) that higher genus Riemann surfaces
 have other degeneration limits- when an open string loop  becomes infinitesimally thin, for example- which may have 
interesting physical interpretations as well. 
\end{itemize}
\noindent We plan to return to some of these questions in the future.

\bigskip
\bigskip
\bigskip

\noindent{\Large \bf Acknowledgements}

\bigskip

This  work was
supported in part by NSF grant PHY-07-57702. The work of M.C. was supported in part by the 2009-10 Siegfried W. Ulmer Dissertation Year Fellowship of UCLA. We  would like to thank  Costas Bachas, John Estes, Jaume Gomis and Darya Krym for useful conversations. M.G. would like to thank the Arnold Sommerfeld Center, LMU Munich for hospitality while this work was completed.


\newpage

\begin{thebibliography}{99}




\bibitem{Maldacena:1997re}
  J.~M.~Maldacena,
  ``The large N limit of superconformal field theories and supergravity,''
  Adv.\ Theor.\ Math.\ Phys.\  {\bf 2} (1998) 231
  [Int.\ J.\ Theor.\ Phys.\  {\bf 38} (1999) 1113]
  [arXiv:hep-th/9711200].

\bibitem{Gubser:1998bc}
  S.~S.~Gubser, I.~R.~Klebanov and A.~M.~Polyakov,
  ``Gauge theory correlators from non-critical string theory,''
  Phys.\ Lett.\  B {\bf 428} (1998) 105
  [arXiv:hep-th/9802109].

\bibitem{Witten:1998qj}
  E.~Witten,
  ``Anti-de Sitter space and holography,''
  Adv.\ Theor.\ Math.\ Phys.\  {\bf 2} (1998) 253
  [arXiv:hep-th/9802150].




\bibitem{Aharony:1999ti}
  O.~Aharony, S.~S.~Gubser, J.~M.~Maldacena, H.~Ooguri and Y.~Oz,
  ``Large N field theories, string theory and gravity,''
  Phys.\ Rept.\  {\bf 323} (2000) 183
  [arXiv:hep-th/9905111].

\bibitem{D'Hoker:2002aw}
  E.~D'Hoker and D.~Z.~Freedman,
  ``Supersymmetric gauge theories and the AdS/CFT correspondence,''
  arXiv:hep-th/0201253.


\bibitem{Witten:1997yu}
  E.~Witten,
  ``On the conformal field theory of the Higgs branch,''
  JHEP {\bf 9707} (1997) 003
  [arXiv:hep-th/9707093].


\bibitem{Seiberg:1999xz}
  N.~Seiberg and E.~Witten,
  ``The D1/D5 system and singular CFT,''
  JHEP {\bf 9904} (1999) 017
  [arXiv:hep-th/9903224].


\bibitem{Vafa:1995bm}
  C.~Vafa,
  ``Instantons on D-branes,''
  Nucl.\ Phys.\ B {\bf 463} (1996) 435
  [arXiv:hep-th/9512078].


\bibitem{Dijkgraaf:1998gf}
  R.~Dijkgraaf,
  ``Instanton strings and hyperKaehler geometry,''
  Nucl.\ Phys.\ B {\bf 543}, 545 (1999)
  [arXiv:hep-th/9810210].

\bibitem{Karch:2000gx}
  A.~Karch and L.~Randall,
  ``Open and closed string interpretation of SUSY CFT's on branes with
  boundaries,''
  JHEP {\bf 0106}, 063 (2001)
  [arXiv:hep-th/0105132].


\bibitem{Aharony:2003qf}
  O.~Aharony, O.~DeWolfe, D.~Z.~Freedman and A.~Karch,
  ``Defect conformal field theory and locally localized gravity,''
  JHEP {\bf 0307} (2003) 030
  [arXiv:hep-th/0303249].



\bibitem{Bachas:2002nz}
  C.~Bachas,
  ``Asymptotic symmetries of $AdS_2$ branes,''
  arXiv:hep-th/0205115.

\bibitem{Bachas:2008jd}
  C.~Bachas,
  ``On the Symmetries of Classical String Theory,''
  arXiv:0808.2777 [hep-th].
 
 
\bibitem{Raeymaekers:2006np}
  J.~Raeymaekers and K.~P.~Yogendran,
  ``Supersymmetric D-branes in the D1-D5 background,''
  JHEP {\bf 0612} (2006) 022
  [arXiv:hep-th/0607150].

\bibitem{Yamaguchi:2003ay}
  S.~Yamaguchi,
  ``AdS branes corresponding to superconformal defects,''
  JHEP {\bf 0306} (2003) 002
  [arXiv:hep-th/0305007].


\bibitem{Raju:2007uj}
  S.~Raju,
  ``Counting Giant Gravitons in $AdS_3$,''
  Phys.\ Rev.\  D {\bf 77} (2008) 046012
  [arXiv:0709.1171 [hep-th]].


\bibitem{Mandal:2007ug}
  G.~Mandal, S.~Raju and M.~Smedback,
  ``Supersymmetric Giant Graviton Solutions in $AdS_3$,''
  Phys.\ Rev.\  D {\bf 77} (2008) 046011
  [arXiv:0709.1168 [hep-th]].

\bibitem{Skenderis:2002vf}
  K.~Skenderis and M.~Taylor,
  ``Branes in AdS and pp-wave spacetimes,''
  JHEP {\bf 0206} (2002) 025
  [arXiv:hep-th/0204054].




\bibitem{D'Hoker:2006uv}
  E.~D'Hoker, J.~Estes and M.~Gutperle,
  ``Interface Yang-Mills, supersymmetry, and Janus,''
  Nucl.\ Phys.\  B {\bf 753} (2006) 16
  [arXiv:hep-th/0603013].

\bibitem{D'Hoker:2007xy}
  E.~D'Hoker, J.~Estes and M.~Gutperle,
  ``Exact half-BPS Type IIB interface solutions I: Local solution and
  supersymmetric Janus,''
  JHEP {\bf 0706} (2007) 021
  [arXiv:0705.0022 [hep-th]].


\bibitem{D'Hoker:2007xz}
  E.~D'Hoker, J.~Estes and M.~Gutperle,
  ``Exact half-BPS type IIB interface solutions. II: Flux solutions and
  multi-janus,''
  JHEP {\bf 0706} (2007) 022
  [arXiv:0705.0024 [hep-th]].


\bibitem{D'Hoker:2007fq}
  E.~D'Hoker, J.~Estes and M.~Gutperle,
  ``Gravity duals of half-BPS Wilson loops,''
  JHEP {\bf 0706} (2007) 063
  [arXiv:0705.1004 [hep-th]].

\bibitem{D'Hoker:2008wc}
  E.~D'Hoker, J.~Estes, M.~Gutperle and D.~Krym,
  ``Exact Half-BPS Flux Solutions in M-theory I, Local Solutions,''
  JHEP {\bf 0808} (2008) 028
  [arXiv:0806.0605 [hep-th]].

\bibitem{D'Hoker:2008qm}
  E.~D'Hoker, J.~Estes, M.~Gutperle and D.~Krym,
  ``Exact Half-BPS Flux Solutions in M-theory II: Global solutions asymptotic
  to $AdS_7 \times  S^4$,''
  JHEP {\bf 0812} (2008) 044
  [arXiv:0810.4647 [hep-th]].

\bibitem{D'Hoker:2009my}
  E.~D'Hoker, J.~Estes, M.~Gutperle and D.~Krym,
  ``Exact Half-BPS Flux Solutions in M-theory III: Existence and rigidity of
  global solutions asymptotic to $AdS4 \times S7$,''
  arXiv:0906.0596 [hep-th].

\bibitem{D'Hoker:2009gg}
  E.~D'Hoker, J.~Estes, M.~Gutperle and D.~Krym,
  ``Janus solutions in M-theory,''
  JHEP {\bf 0906}, 018 (2009)
  [arXiv:0904.3313 [hep-th]].

\bibitem{Clark:2005te}
  A.~Clark and A.~Karch,
  ``Super Janus,''
  JHEP {\bf 0510}, 094 (2005)
  [arXiv:hep-th/0506265].

\bibitem{Yamaguchi:2006te}
  S.~Yamaguchi,
  ``Bubbling geometries for half BPS Wilson lines,''
  Int.\ J.\ Mod.\ Phys.\  A {\bf 22} (2007) 1353
  [arXiv:hep-th/0601089].

\bibitem{Gomis:2006sb}
  J.~Gomis and F.~Passerini,
  ``Holographic Wilson loops,''
  JHEP {\bf 0608} (2006) 074
  [arXiv:hep-th/0604007].

\bibitem{Lunin:2006xr}
  O.~Lunin,
  ``On gravitational description of Wilson lines,''
  JHEP {\bf 0606}, 026 (2006)
  [arXiv:hep-th/0604133].


\bibitem{Lunin:2007ab}
  O.~Lunin,
  ``1/2-BPS states in M theory and defects in the dual CFTs,''
  JHEP {\bf 0710}, 014 (2007)
  [arXiv:0704.3442 [hep-th]].

\bibitem{VanProeyen:1986me}
  A.~Van Proeyen,
  ``Superconformal Algebras,''\\
http://www.slac.stanford.edu/spires/find/hep/www?irn=1943812\\
{\it  in  Vancouver 1986, Proceedings, Super Field Theories, 547-555 }


\bibitem{D'Hoker:2008ix}
  E.~D'Hoker, J.~Estes, M.~Gutperle, D.~Krym and P.~Sorba,
  ``Half-BPS supergravity solutions and superalgebras,''
  JHEP {\bf 0812}, 047 (2008)
  [arXiv:0810.1484 [hep-th]].
 



\bibitem{Chiodaroli:2009yw}
  M.~Chiodaroli, M.~Gutperle and D.~Krym,
  ``Half-BPS Solutions locally asymptotic to $AdS_3 \times S^3$ and interface
  conformal field theories,''
  arXiv:0910.0466 [hep-th].


\bibitem{Kumar:2002wc}
  J.~Kumar and A.~Rajaraman,
  ``New supergravity solutions for branes in  $AdS_3 \times  S^3$,''
  Phys.\ Rev.\  D {\bf 67} (2003) 125005
  [arXiv:hep-th/0212145].

\bibitem{Kumar:2003xi}
  J.~Kumar and A.~Rajaraman,
  ``Supergravity solutions for  $AdS_3\times   S^3$ branes,''
  Phys.\ Rev.\  D {\bf 69} (2004) 105023
  [arXiv:hep-th/0310056].

\bibitem{Kumar:2004me}
  J.~Kumar and A.~Rajaraman,
  ``Revisiting D-branes in $AdS_3 \times   S^3$,''
  Phys.\ Rev.\  D {\bf 70} (2004) 105002
  [arXiv:hep-th/0405024].

\bibitem{Bak:2007jm}
  D.~Bak, M.~Gutperle and S.~Hirano,
  ``Three dimensional Janus and time-dependent black holes,''
  JHEP {\bf 0702} (2007) 068
  [arXiv:hep-th/0701108].




\bibitem{D'Hoker:1988ta}
  E.~D'Hoker and D.~H.~Phong,
  ``The Geometry of String Perturbation Theory,''
  Rev.\ Mod.\ Phys.\  {\bf 60}, 917 (1988).

\bibitem{Fay:73}
 J. D. Fay, 
"Theta Functions on Riemann Surfaces," 
Springer-Verlang (1973).

\bibitem{Strominger:1996sh}
  A.~Strominger and C.~Vafa,
  ``Microscopic Origin of the Bekenstein-Hawking Entropy,''
  Phys.\ Lett.\  B {\bf 379} (1996) 99
  [arXiv:hep-th/9601029].

\bibitem{Maldacena:1996ky}
  J.~M.~Maldacena,
  ``Black holes in string theory,''
  arXiv:hep-th/9607235.


\bibitem{Romans:1986er}
  L.~J.~Romans,
  ``Selfduality For Interacting Fields: Covariant Field Equations For
  Six-Dimensional Chiral Supergravities,''
  Nucl.\ Phys.\  B {\bf 276} (1986) 71.

\bibitem{Tanii:1984zk}
  Y.~Tanii,
  ``N=8 Supergravity In Six-Dimensions,''
  Phys.\ Lett.\  B {\bf 145} (1984) 197.




\bibitem{abramowitz}
M.~Abramowitz and I. Stegun, "Handbook of Mathematical Functions", Dover Publications.

\bibitem{Marolf:2000cb}
  D.~Marolf,
  ``Chern-Simons terms and the three notions of charge,''
  arXiv:hep-th/0006117.



\bibitem{Das:1996jy}
  S.~R.~Das and S.~D.~Mathur,
  ``Interactions involving D-branes,''
  Nucl.\ Phys.\  B {\bf 482} (1996) 153
  [arXiv:hep-th/9607149].


\bibitem{Greene:1989ya}
  B.~R.~Greene, A.~D.~Shapere, C.~Vafa and S.~T.~Yau,
  ``Stringy Cosmic Strings And Noncompact Calabi-Yau Manifolds,''
  Nucl.\ Phys.\  B {\bf 337} (1990) 1.

\bibitem{Gibbons:1995vg}
  G.~W.~Gibbons, M.~B.~Green and M.~J.~Perry,
  ``Instantons and Seven-Branes in Type IIB Superstring Theory,''
  Phys.\ Lett.\  B {\bf 370}, 37 (1996)
  [arXiv:hep-th/9511080].

\end{thebibliography}
\end{document}